\documentclass{revtex4-1}
\usepackage{latexsym, color, graphicx, comment}
\usepackage{bm}
\usepackage{amssymb}
\usepackage{amsmath}
\usepackage{hyperref}
\usepackage{subcaption}
\usepackage{footnote}
\usepackage[toc,page]{appendix}
\hypersetup{colorlinks=true,linkcolor=blue}

\newcommand{\vect}[1]{\mbox{\boldmath $#1$}}
\DeclareMathAlphabet{\mathbfsf}{\encodingdefault}{\sfdefault}{bx}{n}

\begin{document}

\title{The Parallel Boundary Condition for Turbulence Simulations in Low Magnetic Shear Devices}

\author{M. F. Martin}
\email[]{mfmartin@umd.edu}
\affiliation{Institute for Research in Electronics and Applied Physics, University of Maryland, College Park, MD, 20742, USA}

\author{M. Landreman}
\affiliation{Institute for Research in Electronics and Applied Physics, University of Maryland, College Park, MD, 20742, USA}

\author{P. Xanthopoulos}
\affiliation{Max Planck Institute for Plasma Physics, Greifswald, Germany}

\author{N. R. Mandell}
\affiliation{Department of Astrophysical Sciences, Princeton University, Princeton, NJ, 08543, USA}

\author{W. Dorland}
\affiliation{Dept of Physics and Institute for Research in Electronics and Applied Physics, University of Maryland, College Park, MD, 20742, USA}

\begin{abstract}

Flux tube simulations of plasma turbulence in stellarators and tokamaks typically employ coordinates which are aligned with the magnetic field lines. 
Anisotropic turbulent fluctuations can be represented in such field-aligned coordinates very efficiently, but the resulting non-trivial boundary conditions involve all three spatial directions, and must be handled with care. 
The standard ``twist-and-shift'' formulation of the boundary conditions [Beer, Cowley, Hammett \textit{Phys. Plasmas} \textbf{2}, 2687 (1995)] was derived assuming axisymmetry and is widely used because it is efficient, as long as the global magnetic shear is not too small. 
A generalization of this formulation is presented, appropriate for studies of non-axisymmetric, stellarator-symmetric configurations, as well as for axisymmetric configurations with small global shear. 
The key idea is to replace the ``twist" of the standard approach (which accounts only for global shear) with the integrated local shear. 
This generalization allows one significantly more freedom when choosing the extent of the simulation domain in each direction, without losing the natural efficiency of field-line-following coordinates. 
It also corrects errors associated with naive application of axisymmetric boundary conditions to non-axisymmetric configurations. Simulations of stellarator turbulence that employ the generalized boundary conditions require much less resolution than simulations that use the (incorrect, axisymmetric) boundary conditions. 
We also demonstrate the surprising result that (at least in some cases) an easily implemented but manifestly incorrect formulation of the boundary conditions does {\it not} change important predicted quantities, such as the turbulent heat flux. 

\end{abstract}


\maketitle

\section{Introduction} \label{intro}

Understanding and predicting turbulent transport in fusion devices remains one of the most pressing issues in moving fusion energy forward. 
In the tokamak community, microinstabilities and core turbulence have been extensively studied using an array of gyrokinetic codes \cite{GS2,GENE,GYRO,GKV}.
However, solving the gyrokinetic equation is a generally expensive endeavor, and the geometric complexities introduced when moving to stellarators result in commensurately more expensive computational studies.
For instance, a full flux surface gyrokinetic simulation that uses field-line-following coordinates and adiabatic electrons in stellarator geometry using the GENE code \cite{GENE} currently requires on the order of 0.1 M CPU hours.
The most cost-effective option, therefore, is to run these codes in a flux tube ($\sim$10-20 times faster), a simulation domain that follows a magnetic field line and is much longer than it is wide.
Such domains use the field-line-following coordinates and boundary conditions originally developed in \cite{Beer} for gyrofluid simulations. The advantages of flux tube simulations and field-line-following coordinates accrue no matter how one chooses to represent the distribution function (as $f(v)$, moments of $f(v)$, with particles, {\it etc.}). 

The combination of field-line-following coordinates and a flux tube domain reduces turbulence simulation runtimes by $10/\rho_*^2 \sim 10^5$ but requires an implementation of periodicity. 
As the coordinates are non-orthogonal and curvilinear, the flux tube domain boundaries are not manifestly periodic. 
For axisymmetric geometries ({\it e.g.,} tokamaks) the ``twist-and-shift" boundary condition \cite{Beer} has been used for decades and can be  expressed in logically Cartesian $(x,y,z)$ coordinates, with $(x,y)$ measuring appropriately normalized distances in the plane locally perpendicular to the magnetic field and $z$ measuring distances along the magnetic field. 
Expressing the flux tube periodicity involves all three directions but is typically expressed as a ``parallel" boundary condition, as the spatial ``twist" of the magnetic field which is accumulated as one moves along a bundle of magnetic field lines is accommodated by ``shifted" alignments of perpendicular-to-the-field-line Fourier modes at either end of the flux tube. The twist-and-shift boundary conditions \cite{Beer,DimitsQB} were designed to unwind the secular twist that arises from strong global magnetic shear, denoted here by $\hat{s}$.
There are two important consequences of any physically correct twist-and-shift boundary condition.
First, each $(k_x,k_y)$ Fourier mode undergoes a shift in $k_x$ (proportional to $k_y$) across the $z$ boundary. (An equivalent condition exists for non-spectral representations.)
Second, the perpendicular aspect ratio of the simulation domain, $L_x/L_y$, is necessarily quantized.
As will be discussed in Section \ref{flux_tube_overview}, all existing expressions of these constraints explicitly 
depend on the global magnetic shear.

Problems arise in devices such as W7-X \cite{W7-X}, which was designed to have rotational transform with minimal radial variation, to avoid low-order rational flux surfaces.
Low global shear designs are not exclusive to stellarators, however, as advanced tokamak scenarios \cite{ITER_hybrid} can have similarly flat $q$ profiles, where $q$ is the safety factor.
In such geometries, the existing expression of the parallel boundary condition is inconvenient because of the intrinsically low global magnetic shear.
In particular, the perpendicular aspect ratio of the simulation domain is inversely proportional to the global shear $(L_x \propto L_y/\hat{s})$. For $\hat{s} \ll 1$, this imposes restrictive resolution requirements (a large number of $x$ grid points) that increase computation time. 
Moreover, naive application of the axisymmetric formulas to non-axisymmetric geometries results in errors.

To address these shortcomings, we have generalized the parallel boundary condition for flux tube simulations.
Our generalization allows for non-axisymmetric geometries and depends on local rather than global magnetic shear. 
In many cases of interest, the local magnetic shear can vary considerably along a flux tube, even when the global shear is weak.
For 3-D equilibria, our approach requires stellarator symmetry (Section \ref{stell_symmetry}).
In the case of axisymmetry, this generalized boundary condition reduces to conventional ``twist-and-shift" when the flux tube ends are separated by an integer number of poloidal turns.

The significant variation of local magnetic shear in low \textit{global} shear geometries presents the opportunity to optimize the flux tube length. For appropriately selected flux tube lengths, it is possible to use periodic parallel boundary conditions, or to preserve continuity in the magnetic drifts across the parallel boundary with a perpendicular aspect ratio of the simulation domain close to unity. The effects of this new boundary condition on the speed and accuracy of microinstability and turbulence simulations are explored here.

This paper is organized as follows:
In Section \ref{flux_tube_overview}, we define the field-line-following coordinate system and how fluctuating quantities are represented within the simulation domain and across its perpendicular boundaries, with Section \ref{conventional} detailing the conventional method of handling the parallel boundary.
Section \ref{new_bc} presents the full derivation of the new, generalized version of the ``twist-and-shift'' boundary condition, including discussions of some of its useful properties. 
Section \ref{optimization} discusses the characteristic behavior of shear in a stellarator flux tube geometry possessing low global shear, and how this behavior can be used to optimize the new boundary condition.
Finally, Section \ref{results} presents numerical studies
of linear instability, secondary instability, and nonlinear turbulence, showing how relevant physical quantities depend on the parallel boundary condition choice.
In the cases tested here, results indicate that even incorrect implementations of the new boundary condition do not affect the ability of a simulation to predict certain important quantities.
A significant computational speedup is also observed when using the new boundary condition, compared to the conventional method.

\section{Flux Tube Simulations} \label{flux_tube_overview}

The microinstabilities that develop into the turbulence responsible for the high levels of transport observed in fusion devices are characteristically highly elongated along the magnetic field relative to their scale lengths perpendicular to the magnetic field.
It is thus natural to introduce field-aligned coordinates \cite{Beer} for toroidal magnetic confinement devices. Such coordinates are readily understood when the magnetic field is expressed in Clebsch form,
\begin{equation}
  \bm{B} = \nabla\psi\times\nabla\alpha.
\end{equation}
Here, $\psi$ and $\alpha$ are constant on magnetic field lines, and so can be used as coordinates in the plane perpendicular to $\bm{B}$. Without loss of generality, we  identify $\psi$ as a  magnetic surface label ({\it e.g.,} toroidal or poloidal flux) and thus think of it as the logically radial coordinate.
The coordinate $\alpha$ is a magnetic field line label.
The third coordinate $z$ measures distances along a magnetic field line. Again without loss of generality, we identify $z$ with the poloidal angle $\theta$.

The minimal simulation domain for a turbulence simulation should not be shorter than the correlation length in any direction. 
Perpendicular correlation lengths $\lambda$ are observed to be on the order of a few ion gyroradii in core plasmas. 
It should therefore be possible to model these fluctuations in a periodic perpendicular domain of size $L_\psi \times L_\alpha$, as long as $L_\alpha/\lambda, L_\psi/\lambda$ are both large enough. We wish to find a minimal domain and so we use a periodic perpendicular domain whose lengths are measured in ion gyroradii.
For any fluctuating quantity $\phi$,
\begin{eqnarray}
  \phi(\psi,\alpha,z,t) = \phi(\psi+L_{\psi},\alpha,z,t) = \phi(\psi,\alpha+L_{\alpha},z,t).
\end{eqnarray}
The small perpendicular extent of the box also means that geometric quantities $(\bm{B},\nabla\psi,\nabla\alpha)$ can be fully characterized by their local values and gradients, approximately independent of $\psi,\alpha$. 

The periodic perpendicular boundary conditions allow one to represent $\phi$ as a Fourier series in these coordinates:
\begin{equation}
  \phi(\psi,\alpha,z,t) = \sum_{m=-\infty}^{\infty}\sum_{n=-\infty}^{\infty}\hat{\phi}_{m,n}(z,t)\;\mathrm{exp}\Big(\frac{2\pi i(\psi-\psi_0)m}{L_{\psi}} + \frac{2\pi i(\alpha-\alpha_0)n}{L_{\alpha}}\Big),
\end{equation}
with the constants $(\psi_0,\alpha_0)$ representing the center of the flux tube in the perpendicular plane.
For both a simplified representation and a means to understand the steps which follow, the fluctuations can also be represented by the wavenumbers $k_{\psi}\equiv 2\pi m/L_{\psi}$ and $k_{\alpha}\equiv 2\pi n/L_{\alpha}$:
\begin{equation}
  \label{eq:fluc_k}
  \phi(\psi,\alpha,z,t) = \sum_{k_{\psi}=-\infty}^{\infty}\sum_{k_{\alpha}=-\infty}^{\infty}\hat{\phi}_{k_{\psi},k_{\alpha}}(z,t)\;\mathrm{exp}\Big(ik_{\psi}\Delta\psi + ik_{\alpha}\Delta\alpha\Big),
\end{equation}
where $\Delta\psi=\psi-\psi_0$ and $\Delta\alpha=\alpha-\alpha_0$. The rest of this paper concerns the conditions imposed at the ends of the domain in the parallel coordinate, $z$, in the context of fluctuations defined as in Eq.~(\ref{eq:fluc_k}). 

\section{The Standard Parallel Boundary Condition} \label{conventional}

The standard parallel boundary condition \cite{Beer} is based on the assumption that turbulent fluctuations should be statistically identical at two locations with the same poloidal angle (but different toroidal angle) in an \textit{axisymmetric} geometry.
It should be clear that this renders the boundary condition formally incorrect when simulating flux tubes in a stellarator, as the geometry is inherently non-axisymmetric.

Quantitatively, this assumption about turbulent fluctuations produces the following constraint on the fluctuating quantity $\phi$:
\begin{equation}
  \label{eq:bc_axi}
  \phi[\psi,\alpha(\theta+2\pi N,\zeta),z(\theta+2\pi N)] = \phi[\psi,\alpha(\theta,\zeta),z(\theta)].
\end{equation}
Here, we take $\theta$ and $\zeta$ to be the poloidal and toroidal angles, respectively, where magnetic field lines are straight in the $(\theta,\zeta)$ plane.
Further, the field line label is taken to be
\begin{equation}
	\alpha=\zeta-q\theta,
    \label{eq:alpha_def}
\end{equation}
where $q=q(\psi)$.
By applying the above constraint to the fluctuation form (\ref{eq:fluc_k}), one can derive a set of conditions that must be satisfied in the simulation, namely:
\begin{equation}
  \begin{gathered}
  \label{eq:old_twist}
  \left[k_{\alpha}\right]_{z=+\pi N} = \left[k_{\alpha}\right]_{z=-\pi N}, \\
      k_{\psi}' \equiv {\left[k_{\psi}\right]}_{z=+\pi N} - \left[k_{\psi}\right]_{z=-\pi N} = 2\pi N\frac{\mathrm{d}q}{\mathrm{d}\psi}k_{\alpha},
  \end{gathered}
\end{equation}
with $N$ being a positive integer.
Thus, by imposing the constraint in (\ref{eq:bc_axi}), there is a required shift in $k_{\psi}$ that a $(k_{\psi},k_{\alpha})$ Fourier mode of $\phi$ must undergo in passing from one end of the domain to the other.
This results in the standard parallel boundary condition on fluctuating quantities in flux tube simulations:
\begin{equation}
	\phi_{k_{\psi}+k_{\psi}',k_{\alpha}}\left[\theta+2\pi N,t\right]C_{k_{\alpha}} = \phi_{k_{\psi},k_{\alpha}}\left[\theta,t\right],
    \label{eq:oldbc}
\end{equation}
where $C_{k_{\alpha}}$ is a phase factor, $|C_{k_{\alpha}}|=1$.
Since we cannot retain an infinite number of modes in a simulation, the shift in $k_{\psi}$, along with the number of modes we choose to evolve, determines the maximum $k_{\perp}$ value able to be resolved. 

At this point, it is appropriate to introduce the coordinates $(x,y)$, which are the standard normalization-dependent code representations of $(\psi,\alpha)$ that have units of length.
The following normalization choices have been used in the steps that follow:

\begin{equation}
	\frac{\mathrm{d}x}{\mathrm{d}\psi_t} = \frac{1}{aB_0\sqrt{s_0}}, \qquad\qquad \frac{\mathrm{d}y}{\mathrm{d}\alpha} = \frac{a\sqrt{s_0}}{q_0}.
	\label{eq:norm_coord}
\end{equation}
where $x\equiv a\sqrt{\psi_t/\psi_{edge}}$, with $\psi_t$ taken to be the toroidal flux and $\psi_{edge}$ is the value of $\psi_t$ at the plasma edge. In the above definitions, $a$ is a constant representing an effective minor radius, $s_0$ is a flux surface label where $s\equiv\psi_t/\psi_{edge}$, and $B_0=2\psi_{edge}/a^2$ is the reference magnetic field.
Using (\ref{eq:norm_coord}), we can rewrite $k_{\psi}'$ in terms of $x$ and $y$ as

\begin{equation}
    k_{x}' = 2\pi N\hat{s}\;k_y,
\end{equation}
with $\hat{s}\equiv\left(x/q)\mathrm{d}q/\mathrm{d}x\right|_{x=x_0}$.
It is also straightforward to show that these conditions impose a quantization on the perpendicular aspect ratio of the domain

\begin{equation}
  \frac{L_x}{L_y} = \frac{J}{2\pi N\left|\hat{s}\right|},
  \label{eq:old_aspect}
\end{equation}
where $J$ is a nonzero integer that can be set in the code to potentially achieve a more desirable aspect ratio. 
These constraints, when applied to stellarator geometries possessing low global shear, become very restrictive with respect to resolution requirements.
For instance, on the $x/a=0.357$ surface of W7-X, where $\hat{s}=-0.019$, (\ref{eq:old_aspect}) corresponds to $L_x/L_y=8.1J$. 
The radial extent of the simulation domain is forced to be large. 
Here, and in stellarator calculations below, we will take $a$ to be the effective minor radius calculated by VMEC \cite{VMEC}.
Good estimates of heat fluxes and other quantities of interest typically require one to resolve fluctuations with wavenumbers extending up to $k_\perp \rho_i \sim 1$. This is expensive in a radially extended domain. In a spectral decomposition, one has to use a correspondingly large number of Fourier modes. In a grid-based discretization, one has to use a large number of grid points.

\section{The New Parallel Boundary Condition(s)} \label{new_bc}

Our generalization necessarily remains consistent with stellarator symmetry, but relaxes the explicit dependence on global magnetic shear in favor of the integrated local magnetic shear.

\subsection{Stellarator Symmetry} \label{stell_symmetry}

A flux tube demonstrating stellarator symmetry has the property that it is unchanged when rotated by $180^{\circ}$ about an appropriate point.
This symmetry implies that the magnitudes of geometric quantities are equivalent at stellarator symmetric locations.
For our purposes, stellarator symmetry can be summarized by three identities,
\begin{equation}
  |B|_{z_+} = |B|_{z_-} \qquad |\nabla\psi|_{z_+} = |\nabla\psi|_{z_-} \qquad |\nabla\alpha|_{z_+} = |\nabla\alpha|_{z_-},
  \label{eq:stell_symm}
\end{equation}
where $z_{\pm}$ indicates the two ends of the flux tube at $(\psi,\pm\theta,\pm\zeta)$.

As long as two stellarator symmetric locations are farther apart than a few correlation lengths in each direction, the fluctuations at those points will thus be indistinguishable on average. (In the absence of stellarator symmetry, the differences in magnetic geometry would not permit this assertion.) We will assert periodicity only at widely separated, stellarator symmetric points.

It is not generally known how to guarantee that a flux-tube domain is long enough, even when it is long compared to the simulation's correlation lengths. In general, for example, there is a flux of free energy along field lines in a turbulent plasma. In a simulation, this free energy flux is a form of dissipation when it is a net exhaust, but a form of noise when the net flow is into the domain. Only by simulating a full flux surface can one resolve this category of uncertainty.

\subsection{Orthonormal coordinates} \label{ortho}

It is convenient to construct orthonormal coordinates $(u,v)$ to describe the plane perpendicular to the magnetic field.
By doing so, we can explicitly capture the local shear information along a flux tube.
In the traditional non-orthogonal coordinates $(\nabla\psi,\nabla\alpha)$, this information manifests itself in a distortion of the perpendicular plane, hiding potentially useful local magnetic shear information.

\begin{figure}[h!]
	\includegraphics[width=1in]{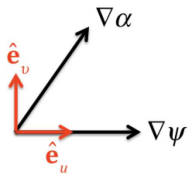}
    \caption{(Color Online) Vector directions in the perpendicular plane for the orthonormal $(\vect{\hat{e}}_u,\vect{\hat{e}}_v)$ and Clebsch $(\nabla\psi,\nabla\alpha)$ coordinates at an arbitrary $z$ location where $\nabla\psi\cdot\nabla\alpha\ne 0$.} 
	\label{fig:orth_coord_fig}
\end{figure}

We consider a Clebsch representation of the magnetic field $\vect{B} = \nabla \psi \times \nabla \alpha$, with a field line centered on the coordinates $\psi ={{\psi }_{0}}$ and $\alpha ={{\alpha }_{0}}$. As before, we assume $\psi$ is a flux surface label. Notice that the following three vectors are orthonormal:
\begin{eqnarray}
\label{eq:unit_vector_def}
\hat{\vect{b}} & = & \frac{\vect{B}}{\left| \vect{B} \right|}, \qquad \hat{\vect{e}}_u = 
\frac{\nabla \psi}{|\nabla\psi|}, \qquad \hat{\vect{e}}_v = \frac{\vect{\hat{b}}\times \nabla \psi}{|\nabla\psi|}.
\end{eqnarray}
We denote the position vector of the central field line by
$\vect{r}_0(z)$, where $z$ parameterizes the position along the field line.
Any point in the flux tube can be labeled with coordinates $(\psi ={{\psi }_{0}}+\Delta \psi, \; \alpha ={{\alpha }_{0}}+\Delta \alpha, \; z)$.
The position vector $\vect{r}$ for this point in the flux tube can be written
\begin{equation}
\vect{r}\left( \psi ,\alpha ,z \right)\approx {{\vect{r}}_{0}}(z)+{{\left( \frac{\partial \vect{r}}{\partial \psi } \right)}_{\alpha ,z}}\Delta \psi +{{\left( \frac{\partial \vect{r}}{\partial \alpha } \right)}_{\psi ,z}}\Delta \alpha.
\label{eq:r}
\end{equation}
At the same time, we can parameterize the perpendicular plane using alternative coordinates $(u,v)$ defined in
terms of the orthonormal basis (\ref{eq:unit_vector_def}):
\begin{eqnarray}
u & = & \left( \vect{r}-{{\vect{r}}_{0}} \right)\cdot \hat{\bm{e}}_u \label{eq:uv_def}, \\ 
v & = & \left( \vect{r}-{{\vect{r}}_{0}} \right)\cdot \hat{\bm{e}}_v. \nonumber
\end{eqnarray}
Substituting (\ref{eq:unit_vector_def}) and (\ref{eq:r}) into (\ref{eq:uv_def}),  noting
\begin{eqnarray}
\left( \frac{\partial \vect{r}}{\partial \psi } \right)_{\alpha ,z}\cdot \nabla \psi = 1, \qquad \left( \frac{\partial \vect{r}}{\partial \psi } \right)_{\alpha ,z}\cdot \nabla \alpha = 0 ,
\end{eqnarray}
and using $\bm{B}=\nabla\psi\times\nabla\alpha$ to find $(\partial\bm{r}/\partial\psi)\cdot\bm{B}\times\nabla\psi = -\nabla\psi\cdot\nabla\alpha$ and $(\partial\bm{r}/\partial\alpha)\cdot\bm{B}\times\nabla\psi = |\nabla\psi|^2$, we obtain a relation between the orthonormal coordinates $(u,v)$ and the standard Clebsch coordinates:
\begin{align}
    u &= \frac{\Delta\psi}{|\nabla\psi|},\nonumber\\
    v &= \frac{-(\nabla\psi\cdot\nabla\alpha)\Delta\psi + |\nabla\psi|^2\Delta\alpha}{B|\nabla\psi|}.
  \label{eq:uv}
\end{align}
Figure \ref{fig:orth_coord_fig} presents an example of how the two sets of coordinates parameterizing the perpendicular plane compare at an arbitrary location along the parallel coordinate, $z$, of the flux tube.

\subsection{Boundary Condition Derivation} \label{derivation}

Using the orthonormal coordinates of (\ref{eq:uv}), the new parallel boundary condition for flux tube simulations can be derived, assuming certain requirements are met. The fluctuations expressed as functions of $\left( u,v \right)$ at the two ends of the flux tube should either be (i) at stellarator-symmetric locations or (ii) separated by an integer number of poloidal turns in axisymmetry.
This allows us to take the fluctuating quantity $\phi$ of (\ref{eq:fluc_k}) to be equal at the two ends of a stellarator-symmetric flux tube, and determine which values of $k_{\psi}$ and $k_{\alpha}$ are connected as in (\ref{eq:old_twist}).

We start by rearranging (\ref{eq:uv}) to get expressions for ($\Delta\psi,\Delta\alpha$) as functions of ($u,v$):
\begin{eqnarray}
\label{eq:transformation}
\Delta\psi & = & u|\nabla\psi|, \\
\Delta\alpha & = & \frac{B}{\left| \nabla \psi  \right|}v+\frac{\nabla \psi \cdot \nabla \alpha }{\left| \nabla \psi  \right|}u. \nonumber
\end{eqnarray}	

Substituting (\ref{eq:transformation}) into (\ref{eq:fluc_k}), we see that the fluctuations for each wavenumber pair $(k_{\psi},k_{\alpha})$ have the form
\begin{equation}
\hat{\phi}_{k_{\psi},k_{\alpha}}(z,t)\exp \left( i|\nabla\psi|\left[ {{k}_{\psi }}+{{k}_{\alpha }}\frac{\nabla \psi \cdot \nabla \alpha }{{{\left| \nabla \psi  \right|}^{2}}} \right]u+i\left[k_{\alpha }\frac{B}{\left| \nabla \psi  \right|}\right]v \right),
\label{eq:fluctuations_uv}
\end{equation}
where terms depending on $u$ have been collected. Identifying the $\left( u,v \right)$ planes at the two ends of the flux tube, then the coefficients multiplying $u$ and $v$ in (\ref{eq:fluctuations_uv}) must each match, yielding:
\begin{equation}
{{\left[ {{k}_{\alpha }}\frac{B}{\left| \nabla \psi  \right|^2} \right]}_{z_+}}={{\left[ {{k}_{\alpha }}\frac{B}{\left| \nabla \psi  \right|^2} \right]}_{z_-}},
\label{eq:kalpha_continuity}
\end{equation}
\begin{equation}
{{\left[ {{k}_{\psi }}+{{k}_{\alpha }}\frac{\nabla \psi \cdot \nabla \alpha }{{{\left| \nabla \psi  \right|}^{2}}} \right]}_{z_+}}={{\left[ {{k}_{\psi }}+{{k}_{\alpha }}\frac{\nabla \psi \cdot \nabla \alpha }{{{\left| \nabla \psi  \right|}^{2}}} \right]}_{z_-}}.
\label{eq:kpsi_continuity}
\end{equation}

These relations hold for all stellarator-symmetric flux tubes, as well as in axisymmetric geometry where the flux tube goes around an integer number of times poloidally, such that the ends coincide in a poloidal projection. 
If neither (i) or (ii) are satisfied, then the magnetic geometry at the two ends of the flux tube is dissimilar and we do not expect the turbulence to be statistically similar at the two ends, so the derivation breaks down.
On the other hand, if either of these two conditions above are satisfied, we can reduce (\ref{eq:kalpha_continuity}) and (\ref{eq:kpsi_continuity}) using (\ref{eq:stell_symm}).
Then (\ref{eq:kalpha_continuity}) indicates that we should link a given ${{k}_{\alpha }}$ to the same ${{k}_{\alpha }}$ at the other end of the flux tube, which is consistent with Beer's result.
We can then write (\ref{eq:kpsi_continuity}) as
\begin{equation}
k_{\psi}^{shift} \equiv [k_{\psi}]_{z_+}-[k_{\psi}]_{z_-} = 
-\left( {{\left[ \frac{\nabla \psi \cdot \nabla \alpha }{{{\left| \nabla \psi  \right|}^{2}}} \right]}_{z_+}}-{{\left[ \frac{\nabla \psi \cdot \nabla \alpha }{{{\left| \nabla \psi  \right|}^{2}}} \right]}_{z_-}} \right){{k}_{\alpha }},
\label{eq:almost_main}
\end{equation}
Stellarator-symmetry allows for further reduction by noting that $\nabla\psi\cdot\nabla\alpha$ is an odd function along the field line, i.e. $[\nabla\psi\cdot\nabla\alpha]_{z_+} = -[\nabla\psi\cdot\nabla\alpha]_{z_-}$:
\begin{equation}
  k_{\psi}^{shift} =\; 2\left(\frac{\left[\nabla\psi\cdot\nabla\alpha\right]_{z_-}}{|\nabla\psi|^2}\right)k_{\alpha} =\; -2\left(\frac{\left[\nabla\psi\cdot\nabla\alpha\right]_{z_+}}{|\nabla\psi|^2}\right)k_{\alpha}
  \label{eq:main_result}
\end{equation}
Equation (\ref{eq:main_result}) is our new boundary condition.
We note here that the quantities $\nabla \psi \cdot \nabla \alpha $ and ${{\left| \nabla \psi  \right|}^{2}}$ determining $k_{\psi}^{shift}$ are already computed in every stellarator gyrokinetic code workflow, as they are needed to compute $k_\perp^2$.
Thus, there are no new geometric quantities that need to be computed in order to use the new boundary condition.
It is also possible to derive the same result if the orthonormal condition is relaxed, and $(\hat{\bm{b}},\hat{\bm{e}}_u,\hat{\bm{e}}_v)$ are taken to be \textit{orthogonal} vectors.

For completeness, using the change of variables employed in (\ref{eq:norm_coord}) for the conventional boundary condition, (\ref{eq:main_result}) can be written in terms of $x$ and $y$ to yield
\begin{equation}
	k_x^{shift} = 2\left(\frac{\left[\nabla x\cdot\nabla y\right]_{z_-}}{|\nabla x|^2}\right)k_y = -2\left(\frac{\left[\nabla x\cdot\nabla y\right]_{z_+}}{|\nabla x|^2}\right)k_y.
    \label{eq:main_xy}
\end{equation}
Finally, one can directly derive a quantization condition on the aspect ratio of the simulation domain from (\ref{eq:main_xy}) to be
\begin{equation}
  \frac{L_x}{L_y} = \frac{J}{2}\frac{|\nabla x|^2}{|\nabla x\cdot\nabla y|},
  \label{eq:aspect_new}
\end{equation}
where $J$ is a nonzero integer.
\subsection{Perpendicular Wavenumber Continuity} \label{kperp_cont}

\begin{figure}[h!]
\includegraphics[width=6in]{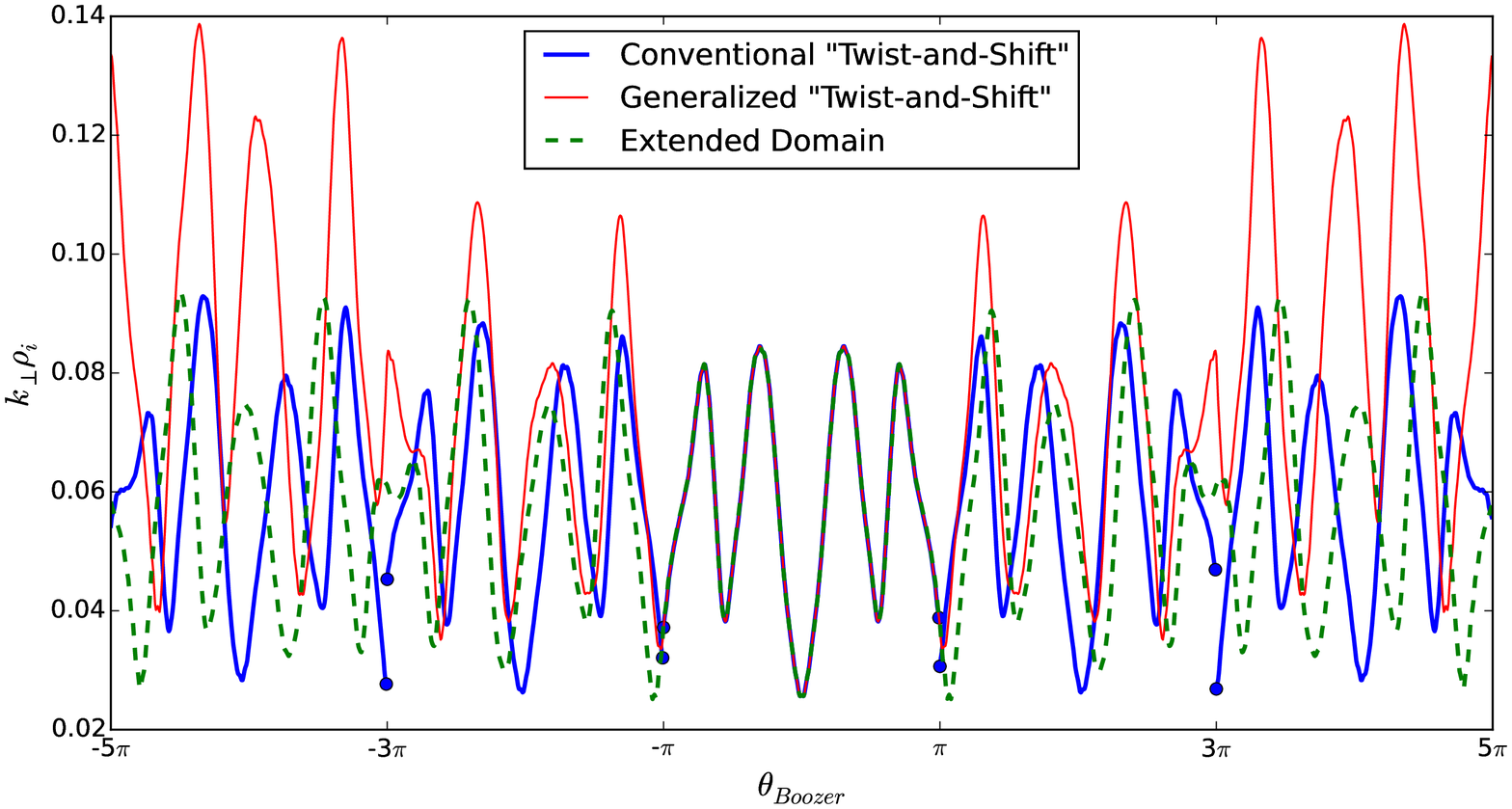}
\caption{(Color online)
The new boundary condition ensures that $k_{\perp}$ remains continuous across
linked domains, whereas the condition of \cite{Beer} generally leads to discontinuities at the boundaries
in a stellarator.
Shown here are five linked domains of parallel length $\Delta \theta = 2\pi$, so the boundaries are at 
 $\theta = \pm \pi$ and $\pm 3\pi$. The calculation here is for the $\alpha=0$ field line on the $x/a$=0.357 surface in W7-X,
considering $k_x\rho_i=0$ and $k_y\rho_i = 0.05$ in the central domain.
(For other choices of $k_y\rho_i$, the curves in the figure would merely be scaled by a constant.)
\label{fig:kperp_continuity}}
\end{figure}

Our formulation manifestly produces perpendicular wavenumbers
\begin{equation}
k_{\perp} = \left( k_\alpha^2 |\nabla\alpha|^2 + 2k_\alpha k_\psi \nabla\alpha\cdot\nabla\psi + k_\psi^2|\nabla\psi|^2 \right)^{1/2},
\label{eq:kperp2}
\end{equation}
that are continuous when passing through the boundary.
In contrast, when the conventional ``twist-and-shift'' condition is used, $k_{\perp}$ is continuous only in the case of axisymmetry with an integer number of poloidal turns.
For the $\alpha=0$ flux tube in W7-X running from $[-\pi,\pi]$, Figure \ref{fig:kperp_continuity} shows a plot of $k_{\perp}$ over a connected domain by linking the flux tube to itself at the boundaries $\pm\pi,\pm 3\pi$ using the boundary conditions in question.
If the conventional boundary condition \cite{Beer} is applied instead of (\ref{eq:main_result}),
then $k_\perp$ becomes discontinuous at the boundary (as one can see in the blue curve) which may cause undesirable numerical behavior.
For Figure \ref{fig:kperp_continuity} and results that follow, we have chosen to normalize wavenumbers to the ion gyroradius, defined to be $\rho_i\equiv v_{ti}/\Omega$, where $v_{ti}\equiv\sqrt{T_i/m_i}$ is the ion thermal velocity, and $\Omega$ is the ion cyclotron frequency.

This continuity might be important because $k_{\perp}$ appears in the argument of the Bessel functions in the gyrokinetic equation, making it noteworthy that $k_\perp$ increases faster with $|\theta|$ for the new boundary condition than for the old condition.
This behavior is expected, since in the old approach, $k_\psi$ is increased by an amount proportional to the small global shear, while in the new approach, $k_\psi$ increases by an amount related to the local shear, which is generally larger.
A large rise in $k_\perp$ with $|\theta|$ is desirable because it leads to localization of the eigenfunctions and turbulence within a small number of linked domains (since the Bessel functions cause the plasma response to decrease with $k_\perp$), leading to less expensive simulations.

Alongside the plots of $k_{\perp}$ using the two boundary conditions in Figure \ref{fig:kperp_continuity}, we have also plotted $k_{\perp}$ in what we call the `extended domain',
meaning a very long flux tube with no
linkages across the tube ends. 
(For this figure, the extended domain represents a tube of length $\ge 10\pi$.)
While the extended domain represents the true magnetic geometry, its length makes nonlinear simulations impractical, so the workaround is to use shorter domains that can be connected.
The behavior of $k_{\perp}$ past the first connection in a linked domain will generally be different than in the extended domain, regardless of the boundary condition choice.

\subsection{Axisymmetric Limit} \label{axi_limit}

Let us now show that (\ref{eq:main_result}) reduces to Beer's condition in axisymmetric geometry if the flux tube extends an integer number of times poloidally around the torus. By using the definition of $\alpha$ in (\ref{eq:alpha_def}), we can write
\begin{equation}
\nabla \psi \cdot \nabla \alpha =\nabla \psi \cdot \nabla \zeta -q\nabla \psi \cdot \nabla \theta -\theta \frac{\mathrm{d}q}{\mathrm{d}\psi}{{\left| \nabla \psi  \right|}^{2}}.
\label{eq:grad_psi_dot_grad_alpha}
\end{equation}
Due to axisymmetry, $\nabla \psi \cdot \nabla \zeta $ is the same at the forward and backward end of the flux tube. The same is true of $\nabla \psi \cdot \nabla \theta $. Therefore, these terms cancel when (\ref{eq:grad_psi_dot_grad_alpha}) is substituted into
(\ref{eq:almost_main}). The remaining term gives
\begin{gather}
  \left(k_{\psi}^{shift}\right)_{AS} \equiv [k_{\psi}]_{+\pi N} - [k_{\psi}]_{-\pi N} = \frac{\mathrm{d}q}{\mathrm{d}\psi}[\theta (z_+)-\theta(z_-)]k_{\alpha},\\
  \left(k_{\psi}^{shift}\right)_{AS} = k_{\psi}' = 2\pi N\frac{\mathrm{d}q}{\mathrm{d}\psi}k_{\alpha}, 
  \label{eq:axisymmetry}
\end{gather}
where $N$ is the number of times the flux tube extends poloidally around the torus.
This is equivalent to the conventional ``twist-and-shift'' boundary condition in (\ref{eq:old_twist}).

Continuity of $k_\perp$ can be shown in axisymmetry
by noting $\nabla\psi$ is the same at $z_{+}$ and $z_-$, and from (\ref{eq:alpha_def}),
\begin{equation}
\left[ \nabla\alpha \right]_{z_+} = \left[ \nabla\alpha \right]_{z_-} - 2\pi N \frac{dq}{d\psi}\nabla\psi.
\end{equation}

\section{Selecting the Flux Tube Length Using Local Magnetic Shear} \label{optimization}

In a stellarator-symmetric flux tube, $\nabla \psi \cdot \nabla \alpha $ is an odd function of $z$.
This can be seen from the fact that $\nabla \psi \cdot \nabla \alpha $ flips sign under the replacements $\left( \theta \to -\theta ,\; \zeta \to -\zeta  \right)$, where now $\theta $ and $\zeta $ are any straight-field-line coordinates satisfying $\alpha =\zeta -q\theta $.
This is why all terms in (\ref{eq:grad_psi_dot_grad_alpha}) generally add and allowed for the last step in producing (\ref{eq:main_result}).
In particular, in a stellarator it is generally not valid to drop the $\nabla \psi \cdot \nabla \zeta $ and $\nabla \psi \cdot \nabla \theta $ terms, even if the flux tube goes an integer number of times around the torus poloidally.

As discussed in \cite{helander_14}, the local magnetic shear is

\begin{equation}
S=\bm{B}\cdot\nabla\left(\frac{\nabla\psi\cdot\nabla\alpha}{|\nabla\psi|^2}\right).
\end{equation}
Therefore, the shift to $k_{\psi}$ in (\ref{eq:main_result}) represents the integral of the local shear along the flux tube, which makes this new boundary condition advantageous for a couple of reasons.
First, $k_{\psi}^{shift}$ is no longer solely dependent on a potentially restrictive constant global shear but rather a locally varying function.
It is important to note, however, that while $\hat{s}$ is not explicit in (\ref{eq:main_result}), the global shear information is contained in the geometric quantities $\nabla\psi\cdot\nabla\alpha$ and $|\nabla\psi|^2$. 
Second, the fact that $k_{\psi}^{shift}$ depends on a function of $z$ evaluated at flux tube ends means that the length of the tube can be chosen such that an optimal $k_{\psi}^{shift}$ is obtained.

\begin{figure}[h!]
\begin{subfigure}[t]{0.45\textwidth}
\includegraphics[width=3in]{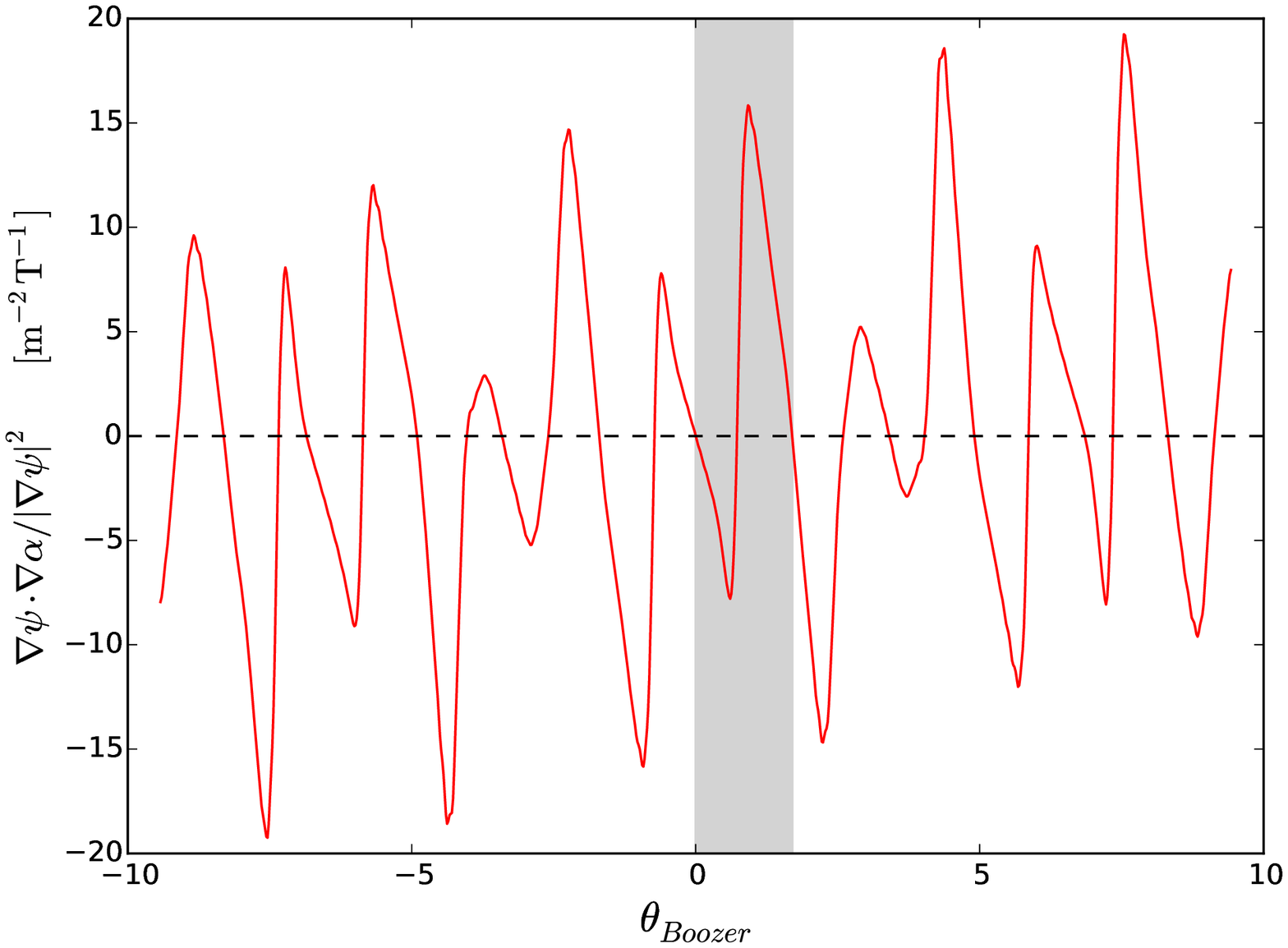}
\caption{}
\label{fig:grad_psi_dot_grad_alpha}
\end{subfigure}
\begin{subfigure}[t]{0.45\textwidth}
\includegraphics[width=3in]{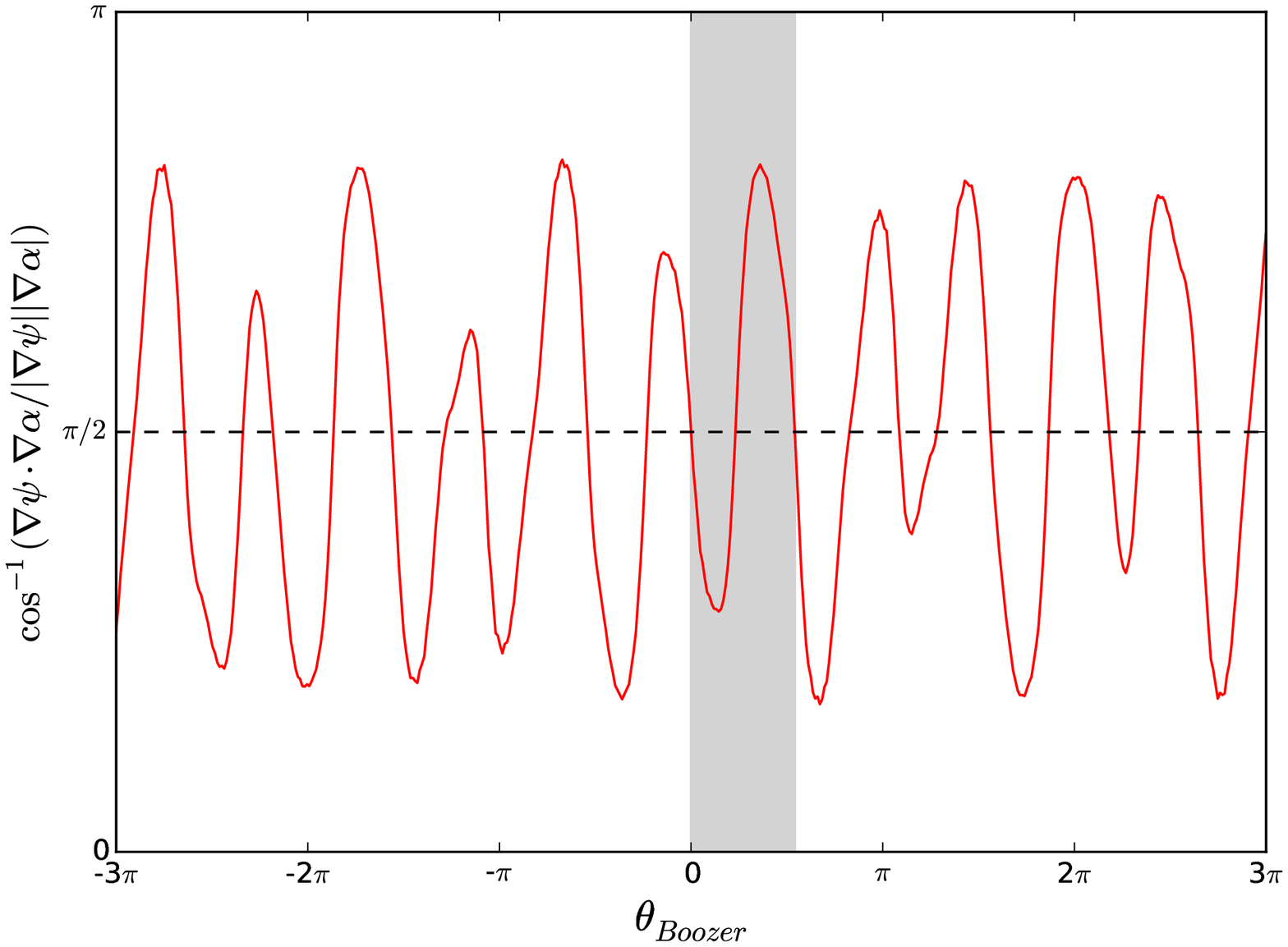}
\caption{}
\label{fig:grad_psi_grad_alpha_angle}
\end{subfigure}
\caption{(Color online)
  (a) Variation of $\nabla\psi\cdot\nabla\alpha / |\nabla\psi|^2$,
the quantity arising in the new boundary condition, in the W7-X standard configuration for the surface with normalized radius $x/a$=0.357 and the field line $\alpha=0$.
  (b) The angle between the two field-aligned coordinate directions $\nabla\psi$ and $\nabla\alpha$, along the same field line.
The large departures from orthogonality $(\pi/2)$ indicate that the local shear is significant even though the global shear is small, $\hat{s} = -0.019$.
Here we take $2\pi\psi$ to be the poloidal flux.
The shaded regions in (a) and (b) indicate the length of the tube in Figure \ref{fig:stell_cross}.}
\label{fig:both_gpsi_galpha}
\end{figure}

Figure \ref{fig:both_gpsi_galpha} gives
some insight into how the local shear and the simulation domain length are related.
In Figure \ref{fig:grad_psi_dot_grad_alpha} we have plotted the integrated local shear, which defines $k_{\psi}^{shift}$ up to a constant.
This curve shows that the local shear has an oscillatory form in this geometry and in fact changes sign a number of times over this domain.
These frequent sign changes in the local shear, and by definition $k_{\psi}^{shift}$, provide the opportunity to make $k_{\psi}^{shift}=0$ if the flux tube length is chosen such that the ends lie where the local shear vanishes.
If $k_{\psi}^{shift}$ vanishes this implies that $[k_{\psi}]_{z_+}=[k_{\psi}]_{z_-}$, which in combination with (\ref{eq:kalpha_continuity}) assures that the parallel boundary condition becomes periodic.
Along with improving computational efficiency, periodic boundary conditions remove the quantization on the aspect ratio of the simulation domain (\ref{eq:aspect_new}).
Now, while in principle one could decide to minimize the length of the tube with this condition in mind by choosing flux tube ends to lie at the first zero of the local shear ($\sim\pi/3$ in this case), this boundary condition only allows for periodicity and \textit{does not imply correct results for an arbitrarily small flux tube}, as will be discussed in the following sections.

\begin{figure}[h!]
\includegraphics[width=3in]{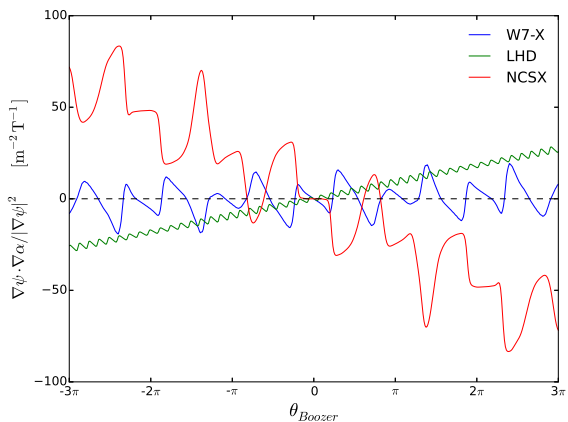}
\caption{(Color online) The quantity $\nabla\psi\cdot\nabla\alpha/|\nabla\psi|^2$ over the domain $[-3\pi,3\pi]$ for the standard equilibrium configurations of W7-X, NCSX, and LHD.
  Each curve denotes the $\alpha=0$ field line at a radial position of $x/a\approx0.36$.
\label{fig:quant_device}}
\end{figure}

This type of behavior that allows for periodic boundary conditions is a result of the low global shear, which permits the oscillations about zero to dominate the functional form of the integrated local shear.
While small, the effect of the global shear is visible in the slight linear trend of the function.
Conversely, the linear trend for a geometry with significant global shear would dominate, and reduce (or perhaps eliminate) any zero crossings in the integrated local shear.
The zero crossings that remain, if any, would then be concentrated near the center and periodic boundary conditions would be limited to shorter flux tubes.
Figure \ref{fig:quant_device} examines this effect by comparing the curve from Figure \ref{fig:grad_psi_dot_grad_alpha} to the same quantity for larger global shear devices, namely LHD \cite{LHD} and NCSX \cite{NCSX}.
These complications don't preclude one from using the boundary condition in high global shear geometries, but the value of optimizing the tube length is more limited.

\begin{figure}[h!]
\includegraphics[width=3in]{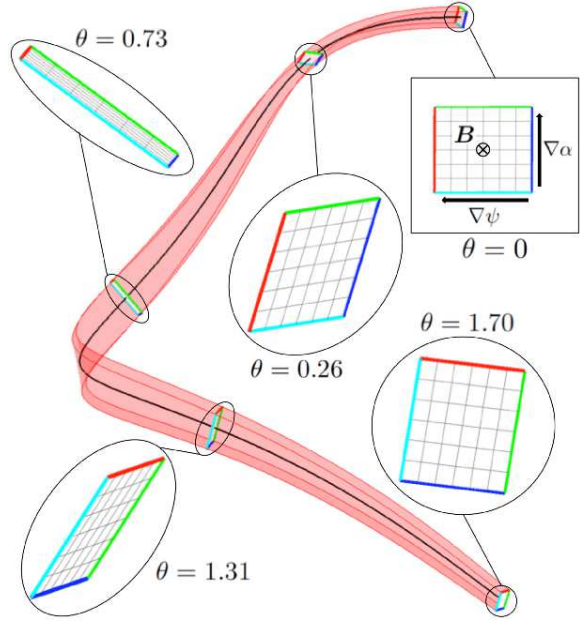}
\caption{(Color online) 3D visualization of the $\alpha=0$ W7-X flux tube in the field-line-following coordinates $(\psi,\alpha,z)$ at $x/a=0.357$.
  The relevant field line is centered at $(\psi_0,\alpha_0)$ in the domain (black line), with the perpendicular boundaries located at $\psi=\psi_0\pm\Delta\psi$ and $\alpha=\alpha_0\pm\Delta\alpha$.
  The color scheme is as follows: $\psi=\psi_0+\Delta\psi$ (red line); $\psi=\psi_0-\Delta\psi$ (blue line); $\alpha=\alpha_0+\Delta\alpha$ (green line); $\alpha=\alpha_0-\Delta\alpha$ (cyan line).
  All cross sections are projected along the magnetic field at the given $\theta$ location.
  The $\theta=0,0.73,1.70$ positions correspond to a vanishing of the integrated local shear (see Figure \ref{fig:grad_psi_dot_grad_alpha}), resulting in a rectangular cross section.}
\label{fig:stell_cross}
\end{figure}

\begin{figure}[h!]
\includegraphics[width=4.5in]{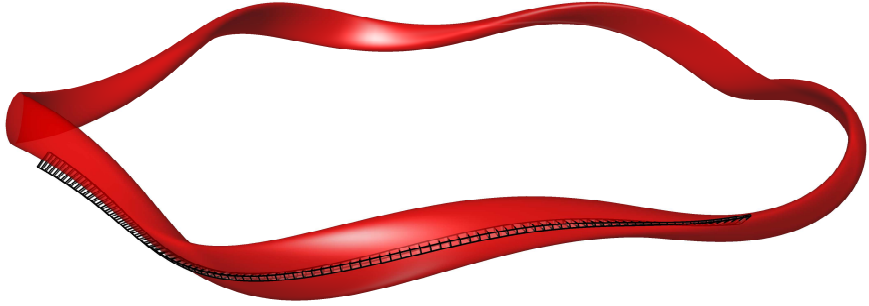}
\caption{(Color online) 3D visualization of the $\alpha=0$ flux tube domain in real space superimposed on the flux \textit{surface} at $x/a=0.357$. (The extent of the tube in $\psi$,$\alpha$ was set for visualization purposes). This is the same flux tube from Figure \ref{fig:stell_cross}, but shown from $\theta=[-0.73,0.73]$, where $\nabla\psi\cdot\nabla\alpha=0$ and the ends of the tube have a rectangular perpendicular cross section.}
\label{fig:tube_surface}
\end{figure}

To get more of a sense for what is happening physically, Figures \ref{fig:stell_cross} and \ref{fig:tube_surface} illustrate how the local shear influences the overall shape of a flux tube.
Figure \ref{fig:stell_cross} shows the $\alpha=0$ W7-X flux tube at $x/a=0.357$ extending from the outboard midplane at $\theta=0$ (bean cross section) to $\theta=1.70$ and is meant to coincide with the shaded regions of Figure \ref{fig:grad_psi_dot_grad_alpha}.
The difference between high and low global shear cases is clear as the 3-dimensional shape of the flux tube constitutes a twisting-and-untwisting of the domain, in contrast to the near monotonic twisting of high global shear flux tube.
Since the condition for periodic parallel boundary conditions require $\nabla\psi\cdot\nabla\alpha=0$ (implying a rectangular perpendicular cross section), this twisting-untwisting characteristic of the flux tube is what affords many potential lengths for which periodic boundary conditions are possible.

\subsection{Magnetic Drift Continuity} \label{drifts}

The magnetic drift term in the gyrokinetic equation (Appendix \ref{gk_eqn}), $\bm{v}_m\cdot\nabla_{\perp}h$, is continuous in axisymmetry with the standard twist-and-shift condition, but the term is generally discontinuous across the parallel boundary of a flux tube in a stellarator, for both the old and new boundary conditions.
It is not obvious that continuity of this term matters, since discontinuity of coefficients in a PDE does not necessarily cause the solution to be discontinuous or otherwise pathological.
To investigate whether it makes a difference, it is possible to make the magnetic drift term continuous in the steps that follow.
We begin by taking the $\nabla B$-drift part of the magnetic drift term in the gyrokinetic equation (\ref{eq:gk_eqn}), noting that $\nabla_{\perp}=i(k_{\psi}\nabla\psi + k_{\alpha}\nabla\alpha)$:
\begin{gather}
  \bm{v}_m\cdot\nabla_{\perp}h \propto \bm{B}\times\nabla B\cdot(k_{\psi}\nabla\psi + k_{\alpha}\nabla\alpha)h.
\end{gather}
Setting $\bm{v}_m\cdot\nabla_{\perp}h$ equal at both ends of the tube, we note that the $\nabla\alpha$- and $\nabla\psi$-components of the $\nabla B$-drift are even and odd functions, respectively, in stellarator-symmetric flux tubes.
Applying this fact along with (\ref{eq:main_result}), the resulting condition can be derived:
\begin{gather}
	\left(\left[k_{\psi}\right]_{z_+} + \left[\frac{\nabla\psi\cdot\nabla\alpha}{|\nabla\psi|^2}\right]_{z_+}k_{\alpha}\right)\;\left[\bm{B}\times\nabla B\cdot\nabla\psi\right]_{z_+} = 0.
\end{gather}
This condition can only be satisfied for all $k_\psi$ and $k_\alpha$ in the simulation when $[\bm{B}\times\nabla B\cdot\nabla\psi]_{z_+}=0$.
In other words, the only way to make the magnetic drifts  continuous across the parallel boundary is to choose the flux tube length such that the radial component of the $\nabla B$-drift vanishes at the ends.
The argument here is identical for the curvature drift as well, since $\bm{B}\times\nabla B\cdot\nabla\psi=0$ is equivalent to $\bm{B}\times\bm{\kappa}\cdot\nabla\psi=0$, at any $\beta$ (normalized pressure).
Therefore, both the $\nabla B$ and curvature drifts become continuous at the same tube length.

\begin{figure}[h!]
\includegraphics[width=3in]{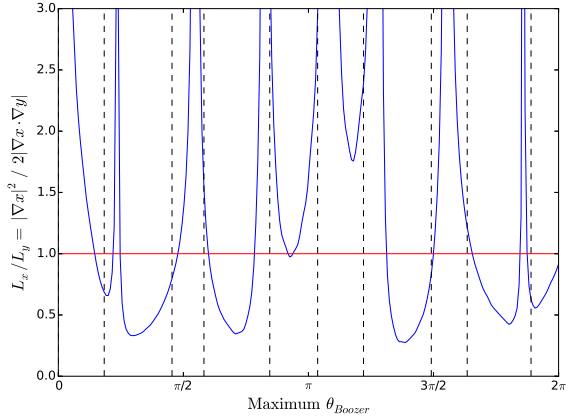}
\caption{(Color online) The solid blue curve shows the quantized domain aspect ratio  as a function of the flux tube's maximum $\theta_{\mathrm{Boozer}}$, for $J=1$. (The flux tube length is twice the maximum $\theta_{Boozer}$.) The horizontal red line represents the ideal $L_x/L_y=1$ case. The dashed vertical lines correspond to flux tube lengths for which $\bm{B}\times\nabla B\cdot\nabla\psi=0$,
so the magnetic drift term is continuous.}
\label{fig:aspect_gb}
\end{figure}

Similar to the quantity $\nabla\psi\cdot\nabla\alpha$ discussed above, $\bm{B}\times\nabla B\cdot\nabla\psi$ varies significantly along a field line, and in fact has many zero-crossings regardless of the global magnetic shear, which is clear from its form in the s-alpha model, $\left(\bm{B}\times\nabla B\cdot\nabla\psi\right)_{s-\alpha} \approx (2a/R)\,\hat{s}\sin(z)$.
Unfortunately, locations where $\nabla\psi\cdot\nabla\alpha$ and $\bm{B}\times\nabla B\cdot\nabla\psi$ vanish do not coincide, meaning magnetic drift continuity cannot be accompanied by appropriately enforced periodic boundary conditions.
However, as one can see in Figure \ref{fig:aspect_gb}, there are numerous locations along a field line where the aspect ratio quantization condition (solid blue curve) approaches unity at the same time as $\bm{B}\times\nabla B\cdot\nabla\psi=0$ occurs (vertical dashed lines). The solid blue curve demonstrates how the length of the flux tube affects the required aspect ratio, with the vertical dashed lines representing locations where $\bm{B}\times\nabla B\cdot\nabla\psi=0$.
The effects of magnetic drift continuity in simulations are explored in the following sections.

\section{Numerical Results} \label{results}

Many questions related to the boundary condition are generic with respect the representation of the distribution function.
The majority of simulations we present below used the GPU-based gyrofluid code GryfX \cite{GryfX}. While GryfX has the option to employ a hybrid approach to simulate zonal flow dynamics with a gyrokinetic model, we have chosen to use GryfX in a pure gyrofluid configuration, in which all modes are evolved using the 4+2 set of gyrofluid equations in \cite{gyrofluid}.
Compared to any comprehensive gyrokinetic model, a gyrofluid model is very inexpensive.
The speedup that is achieved by using GryfX allows for simulations with extremely large $N_x$ (the number of grid points in the $x$ direction) values which would otherwise be computationally impractical in gyrokinetics. This in turn facilitates a more complete survey of the boundary condition issues.

\subsection{Linear Convergence Results} \label{linear}

We consider linear problems first.
Linear flux tube stability analyses have been performed in W7-X geometry in the collisionless, electrostatic, adiabatic electron limit. All simulations use $\alpha=0$ (bean cross section) flux tubes, with geometric information 
calculated by applying the GIST code \cite{GIST} to a VMEC equilibrium. Each flux tube is located at the radial position $x/a=0.357$,
and unless otherwise specified, references to simulated perpendicular wavenumbers are normalized to $\rho_i$. 

\subsubsection{Growth Rate Convergence} \label{growth_rate}

We used the gyrofluid code GryfX \cite{GryfX} to investigate growth rate convergence with respect to both the number of simulated radial modes and length of the flux tube for various boundary condition choices. We assume $T_i/T_e=1$ and equilibrium scale lengths of $a/L_T=3.0$ and $a/L_n=0.0$. One
boundary condition considered is the conventional ``Twist-and-Shift'' condition; in this case the flux tube length is taken to be exactly an integer or half-integer number of poloidal turns. A second option is the new boundary condition, applied to a flux tube with length chosen so
$\left[\bm{B}\times\nabla B\cdot\nabla\psi\right]_{z_{\pm}}=0$  (`Continuous Magnetic Drifts').
A third option, which we will call `Exact periodicity', is periodicity in $z$ imposed for a flux tube with length chosen so
$\left[\nabla\psi\cdot\nabla\alpha\right]_{z_{\pm}}=0$, consistent with the new boundary condition.
The fourth option, which we call `Forced Periodicity', is to impose periodicity in $z$ for a tube length at which 
there is no rigorous analytic justification for doing so, since
$\left[\nabla\psi\cdot\nabla\alpha\right]_{z_{\pm}}\ne 0$. In this case we choose the flux tube length to be exactly an integer or half-integer number of poloidal turns.

The first convergence study examines how the growth rates for each boundary condition choice change as a function of $N_x$ for two binormal wavenumbers $k_y=0.2,0.5$.
Each flux tube has been chosen to be $\sim$1 poloidal turn in length.
The $N_x$ we refer to in this paper is defined to be the number of \textit{aliased} radial modes, where the actual number of simulated (dealiased) radial modes is $\sim 2/3\,N_x$.
These simulated radial wavenumbers are integer multiples of the minimum radial wavenumber, defined by $k_{\psi}^{\mathrm{min}}\equiv 1/L_{\psi}$, where the modes are connected via $k_{\psi}^{shift}$.

\begin{figure}[h!]
\includegraphics[width=3in]{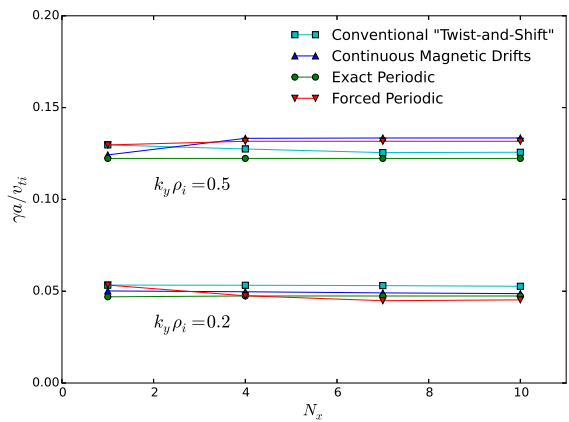}
\caption{(Color online) - Growth rate for $\sim$1 poloidal turn flux tubes as a function of the number of simulated radial modes. The lengths of the flux tubes for each boundary condition choice: Conventional ``Twist-and-Shift"/Forced Periodic $[-\pi,\pi]$, Exact Periodic $[-1.086\pi,1.086\pi]$, Continuous Magnetic Drifts $[-1.045\pi,1.045\pi]$.
}
\label{fig:gamma_nx}
\end{figure}

\begin{figure}[h!]
\includegraphics[width=3in]{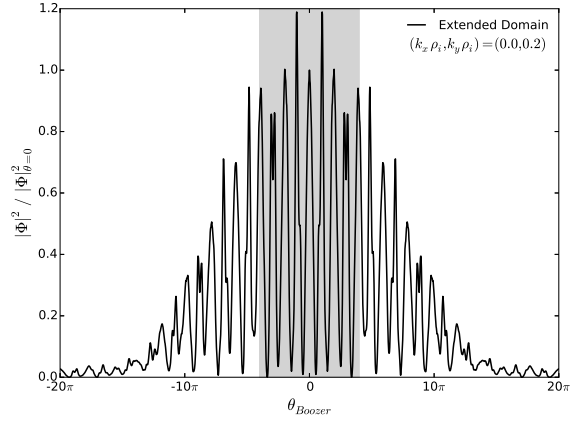}
\caption{Eigenfunction for the $(k_x,k_y)=(0.0,0.2)$ mode over the extended domain $[-20\pi,20\pi]$. The shaded region indicates the extent of the plots in Figure \ref{fig:eig_compare}.
}
\label{fig:extended_section}
\end{figure}

In Figure \ref{fig:gamma_nx}, results make clear that regardless of the chosen boundary condition,  $N_x$ has a very minimal effect on the calculated linear growth rate, and moreover, for $N_x\geq 4$ the growth rate has reasonably converged for both $k_y$ values.
This leads one to think that only a few connected domains (or in some cases only a single $k_x$ value) are necessary to reproduce the extended domain result (i.e. the result in an extremely long flux tube), the eigenfunction of which is displayed in Figure \ref{fig:extended_section} for the $(k_x,k_y)=(0.0,0.2)$ mode.

\begin{figure}[h!]
\includegraphics[width=4in]{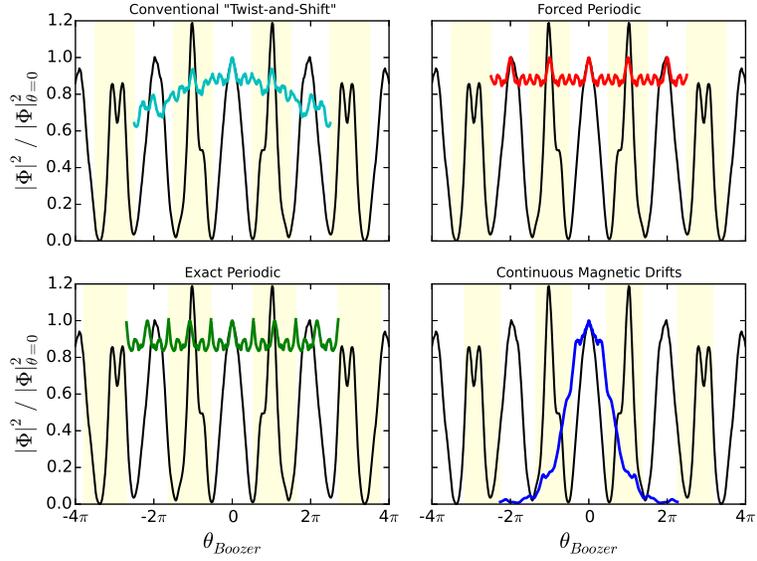}
\caption{(Color Online) Using the 4 boundary conditions from Figure \ref{fig:gamma_nx} for flux tubes of $\sim$0.5 poloidal turn, eigenfunctions for the connected regions ($N_x=7$) are plotted over a portion of the extended domain eigenfunction (black line). The center region in each plot corresponds to the $(k_x,k_y)=(0.0,0.2)$ mode, where the adjacent shaded regions have $(k_x,k_y)=(k_x^{shift},0.2)$, where $k_x^{shift}$ depends on the boundary condition choice.
The lengths of the flux tubes for each boundary condition choice: Conventional ``Twist-and-Shift"/Forced Periodic $[-\pi/2,\pi/2]$, Exact Periodic $[-0.54\pi,0.54\pi]$, Continuous Magnetic Drifts $[-0.45\pi,0.45\pi]$.
}
\label{fig:eig_compare_nx7}
\end{figure}

\begin{figure}[h!]
\includegraphics[width=4in]{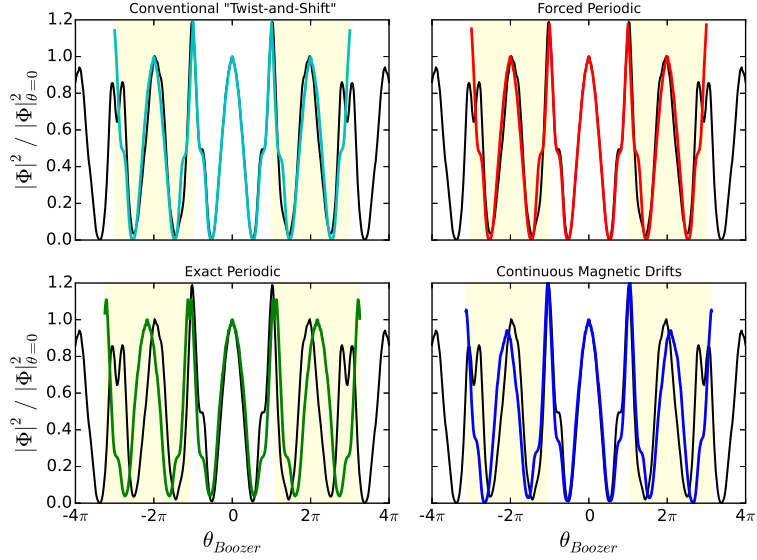}
\caption{(Color online) Using the 4 boundary conditions from Figure \ref{fig:gamma_nx} for flux tubes of $\sim$1 poloidal turn, eigenfunctions for the connected regions ($N_x=4$) are plotted over a portion of the extended domain eigenfunction (black line). 
Exact flux tube lengths are given in the caption of Figure \ref{fig:gamma_nx}.
}
\label{fig:eig_compare}
\end{figure}

In order to better understand how changing $N_x$ and the flux tube length affect our ability to reproduce the extended domain solution, we compare its eigenfunction to the eigenfunctions generated in shorter flux tubes employing the various boundary condition options.
To visualize this, we zoom in on the shaded region of Figure \ref{fig:extended_section} and superimpose plots of eigenfunctions of the connected modes for flux tubes of length $\sim$0.5 poloidal turns $(N_x=7)$ and $\sim$1 poloidal turn $(N_x=4)$ in Figures \ref{fig:eig_compare_nx7} and \ref{fig:eig_compare}, respectively, for each boundary condition choice.

The central region of each plot in Figures \ref{fig:eig_compare_nx7} and \ref{fig:eig_compare} correspond to the mode $(k_x,k_y)=(0.0,0.2)$, which is connected at each end (in the yellow-shaded region) to the eigenfunctions of modes with the same $k_y$ and a different $k_x$ determined by the $k_x^{shift}$ calculated from each boundary condition. As $N_x$ increases, more modes with the same $k_y$ are linked, with $k_x$ changing by integer multiples of $k_x^{shift}$.

By comparing the various boundary conditions (colored lines) to the solid black curve of the extended domain in Figure \ref{fig:eig_compare_nx7} for the $\sim$0.5 poloidal turn flux tube, it is apparent that none of the boundary condition options reliably model the form of the extended domain eigenfunction.
Furthermore, the connected eigenfunction of the Continuous Magnetic Drifts case is distinctly more narrow than the other options.
This seemingly peculiar structure is based on how the $k_x$ dependence of $k_{\perp}$ changes at connection points based on $k_{x}^{shift}$.
For comparatively larger $k_{x}^{shift}$, $k_{\perp}$ will increase accordingly at each connection (see Figure \ref{fig:kperp_continuity}), leading to more localized eigenfunctions.
Moreover, shorter flux tubes will have more connections per unit length, causing this increase in $k_{\perp}$ to occur more frequently, introducing further localization.
This larger $k_{x}^{shift}$ (and slightly shorter flux tube) in the Continuous Magnetic Drifts case is the cause of its comparatively narrow eigenfunction, relative to the smaller shift resulting from the conventional method, and $k_x^{shift}=0$ for the two periodic boundary conditions.
Such disagreement among the boundary condition choices in addition to the poor reconstruction of the extended domain eigenfunction might lead one to expect that growth rate results would be inaccurate, with the continuous magnetic drift result being the biggest outlier.

\begin{figure}[h!]
\includegraphics[width=3in]{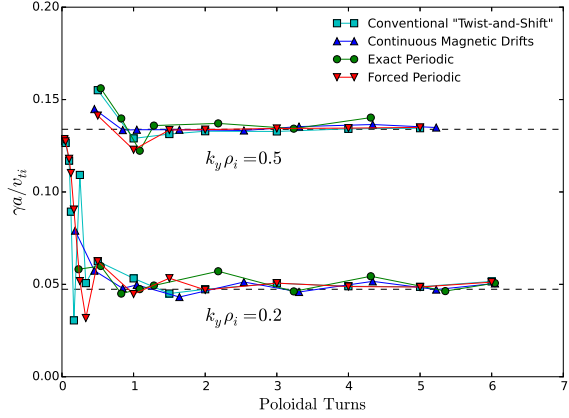}
\caption{(Color online) Linear growth rates as a function of flux tube length for the binormal wavenumbers $k_y=0.2,0.5$, using the boundary conditions described in Section \ref{growth_rate}. The dashed lines represent the \textit{true} results, obtained from the extended domain simulations. In each simulation, the number of $\theta$ grid points, $N_z$, is scaled proportionally with flux tube length, maintaining a fixed $\theta$ resolution for each run.
}
\label{fig:linear_length}
\end{figure}

However, in Figure \ref{fig:linear_length}, which shows the growth rate as a function of flux tube length, the relevant data points at $\sim$0.5 poloidal turns do not support this line of thinking. The growth rates from each boundary condition are closely clustered and are within $\sim30\%$ of the true result.

This same comparison of connected eigenfunctions has been done for flux tubes of $\sim$1 poloidal turn in Figure \ref{fig:eig_compare}.
Unlike $\sim$0.5 poloidal turn flux tubes, agreement with the functional form of the extended domain eigenfunction is quite good, and the boundary condition options have only minor differences between one another.
As one might expect, the growth rate is closer to the true result from Figure \ref{fig:linear_length}, but the extreme contrast between Figures \ref{fig:eig_compare_nx7} and \ref{fig:eig_compare} make it surprising that the growth rates with $\sim$0.5 poloidal turn flux tubes even come close to the true result.
An interpretation of this result is given in Appendix \ref{appendix}.

The  conclusion from the results in this section is that the parallel boundary condition has a seemingly insignificant effect on linear growth rates, as long as the flux tube length exceeds some minimum value.
However, it remains to be shown in Section \ref{nonlinear} how these findings translate to nonlinear simulations when the modes become coupled.

\subsubsection{Linear Zonal Flow Response} \label{rh_residual}

Due to the importance of zonal flows in the saturation of turbulence, representing the response as accurately as possible is advantageous in simulations.
For this reason, understanding the behavior of the dynamic zonal flow response and
Rosenbluth-Hinton (RH) \cite{RH} residual values as the flux tube length is varied is desirable.
It is important to note here that $k_y=0$ modes are self-periodic for both the conventional and generalized twist-and-shift boundary conditions. Hence, the choice of boundary condition affects $k_y=0$ modes only through the tube length, with no effect on the linkages at the parallel boundary.
For calculations in this section, the gyrokinetic code GS2 was used in lieu of GryfX for the purpose of avoiding the closure approximations of the gyrofluid set of equations, which have historically had difficulties in matching zonal flow responses well \cite{Dimits2000}.
The GS2 normalizations are slightly different from GryfX, so for this section we normalize results to the GS2 ion gyroradius, $\rho_{i,\mathrm{GS2}}=\sqrt{2}\rho_i$, and thermal velocity, $v_{ti,GS2}=\sqrt{2}v_{ti}$.

For the figures in this section, the results are produced from flux tube lengths chosen such that the ``Continuous Magnetic Drifts" (blue) and ``Exact Periodic" (red) boundary condition options are applicable, which correspond to  $\left[\mathbf{B}\times\nabla B\cdot\nabla\psi\right]_{z_{\pm}}=0$ and $\left[\nabla\psi\cdot\nabla\alpha\right]_{z_{\pm}}=0$, respectively.
The flux tubes where $\left[\mathbf{B}\times\nabla B\cdot\nabla\psi\right]_{z_{\pm}}=0$ are of particular interest, as linear studies \cite{Monreal,Mishchenko} reveal a dependence on the radial bounce-averaged magnetic drift of the zonal flow residual in stellarators, a quantity that vanishes in axisymmetry.
The bounce-average of $\mathbf{B}\times\nabla B\cdot\nabla\psi$ will thus be performed between two points where this term vanishes, making it possible that such flux tube lengths could result in unique zonal flow behavior compared to other tube lengths.

\begin{figure}[h!]
\includegraphics[width=3in]{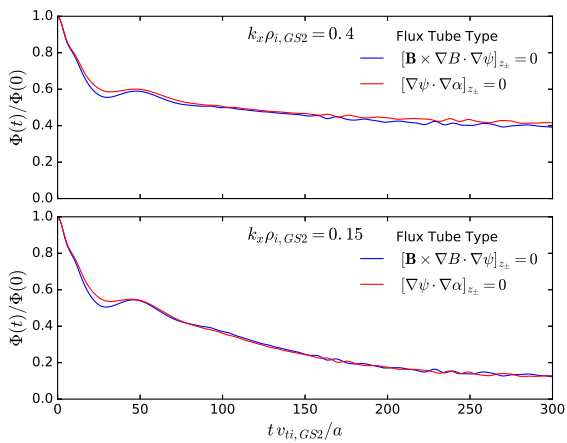}
\caption{(Color online) Gyrokinetic (GS2) simulations of the linear zonal flow response for flux tubes lengths of $\sim$1 poloidal turn for radial wavenumbers $k_x\rho_{i,GS2}=0.15,0.4$, where $\rho_{i,GS2}=\sqrt{2}\rho_i$ and $v_{ti,GS2}=\sqrt{2}v_{ti}$.
The blue and red curves for each radial wavenumber correspond to tube lengths satisfying the conditions $\left[\mathbf{B}\times\nabla B\cdot\nabla\psi\right]_{z_{\pm}}=0$ or $\left[\nabla\psi\cdot\nabla\alpha\right]_{z_{\pm}}=0$, respectively.}
\label{fig:dynamic_zonal}
\end{figure}

Numerical studies have shown that in stellarator geometries, the dynamic response of zonal flows has a central role in the regulation of turbulent transport \cite{Xanth_dynamic}.
In Figure \ref{fig:dynamic_zonal}, this linear response is plotted for $k_x\rho_{i,GS2}=0.15,0.4$ at flux tube lengths of $\sim$1 poloidal turn.
For both wavenumbers in Figure \ref{fig:dynamic_zonal}, the response is nearly identical for both flux tube types. 
This leads to the expectation that the nonlinear effect of the zonal flows will likely be quite similar for both boundary condition options, which is confirmed by the results of Section \ref{nonlinear}.

\begin{figure}[h!]
\includegraphics[width=3in]{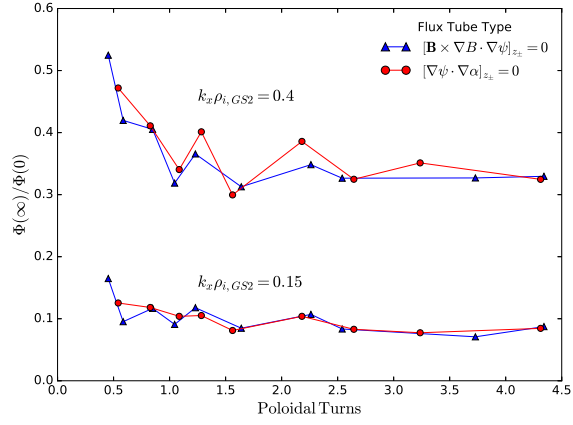}
\caption{(Color online) Gyrokinetic (GS2) calculations of the Rosenbluth-Hinton residual (fit to an exponential decay model) as a function of flux tube length for radial wavenumbers $k_x\rho_{i,\mathrm{GS2}}=0.15,0.4$, where $\rho_{i,\mathrm{GS2}}=\sqrt{2}\rho_i$.
The blue and red curves for each radial wavenumber correspond to tube lengths satisfying the conditions $\left[\mathbf{B}\times\nabla B\cdot\nabla\psi\right]_{z_{\pm}}=0$ or $\left[\nabla\psi\cdot\nabla\alpha\right]_{z_{\pm}}=0$, respectively.}
\label{fig:zonal}
\end{figure}

The long-time zonal flow behavior was also studied, where the RH residual value has been calculated for $k_x\rho_{i,\mathrm{GS2}}=0.15,0.4$ in a variety of flux tube lengths, in Figure \ref{fig:zonal}.
For calculations done with flux tubes less than a full poloidal turn, there is an apparent downward trend in the residual for both flux tube types as the length is increased.
For longer flux tubes the residual values have an oscillatory behavior with an amplitude approaching some constant value as the length is increased.
The results of Figure \ref{fig:zonal} demonstrate that although $\left[\mathbf{B}\times\nabla B\cdot\nabla\psi\right]_{z_{\pm}}=0$ in the blue curve, it appears to have a minor effect relative to flux tubes where this quantity does not vanish at the ends of the domain.

\subsection{Secondary Instability} \label{secondary}

We demonstrate in this section that there is a case,  the evolution of a secondary instability, in which the  discontinuity associated with an incorrectly applied boundary condition could have an effect on results.

The nonlinear generation of zonal flows in plasmas are due in part to a Kelvin-Helmholtz-like secondary instability \cite{Rogers-2000} that develops from the primary Ion-Temperature-Gradient (ITG)-driven radial streamers.
The primary ITG instability (characterized by $k_x=0,k_y\ne 0$) is nonlinearly coupled to a $k_x\ne0,k_y=0$ mode through a three-wave interaction with another unstable ($k_x\ne0,k_y\ne0$) mode, sometimes referred to as the pump wave.

\begin{figure}[h!]
\includegraphics[width=6in]{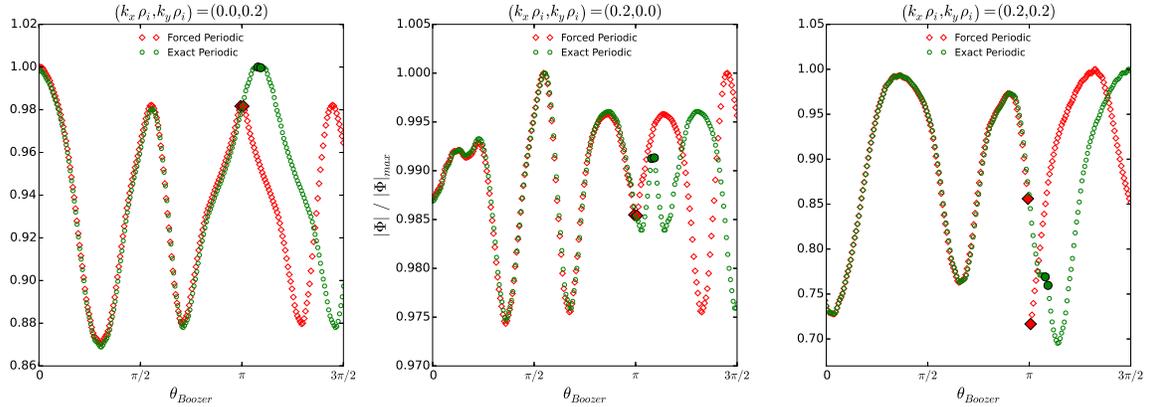}
\caption{(Color online) Eigenfunctions for the primary mode (left), $k_y=0$ mode (center), and pump wave (right) in $\sim$1 poloidal turn flux tubes. The larger, filled markers denote grid points at the ends of the domain, for the ``Exact" (circles) and ``Forced" (diamonds) periodic cases. Note the suppressed zeros and different vertical axes in each figure.}
\label{fig:secondary}
\end{figure}

Returning to GryfX simulations, we address the behavior of the aforementioned modes involved in the generation of the secondary instability when subjected to flux tubes of $\sim$1 poloidal turn, employing the ``Exact" and ``Forced" periodic boundary conditions. 
For the simulation performed here, we begin with a short linear setup run to initialize the primary mode.
The simulation is then nonlinearly restarted with a primary mode amplitude so large that the nonlinear term dominates the equation, allowing one to study the interactions among only the three aforementioned modes.  
The resulting eigenfunctions are plotted in Figure \ref{fig:secondary}, using the finite wavenumbers $k_x=k_y=0.2$.

One can see from the $k_y=0$ (center) and primary mode (left) eigenfunctions that the particular boundary condition does not affect the continuity of these modes across the connected domain.
However, the boundary condition appears to have a notable effect on the pump wave (right) in the form of a discontinuity in the eigenfunction across the connections at $\theta_{Boozer}\approx\pm\pi$.
Such behavior arises through the discontinuity in $k_{\perp}$ between connected domains when using a boundary condition option that is not strictly valid, such as incorrectly enforcing periodicity as we have done here.

This discontinuity in Figure \ref{fig:secondary}.c can be understood by considering the dominant terms in the gyrokinetic equation (Appendix \ref{gk_eqn}) with $k$ corresponding to the pump wave:

\begin{equation}
	\frac{\partial h_k}{\partial t} + \sum_{k',k^{''}}\frac{c}{B}\left\{\langle\phi_{k'}\rangle,h_{k^{''}}\right\}+\dots\quad,
\end{equation}
where the angled brackets $\left\langle...\right\rangle$ denote a gyroaveraging operation performed at constant  guiding center, and $\{\cdot,\cdot\}$ is the Poisson bracket.
For simulations using forced periodicity, even if $\phi_{k'}$ and $h_{k''}$ are continuous, the discontinuity in $k_{\perp}$ will cause the Bessel function $J_0(k_{\perp}\rho_i)$ involved in $\left\langle...\right\rangle$ to be discontinuous, causing $h_k$ to develop a discontinuity.
Thus, incorrectly enforcing periodicity, or otherwise improperly using conventional ``twist-and-shift" in the parallel direction introduces errors.
However, the consequences resulting from introducing these discontinuities are not well understood, and further study is warranted to quantify the full effect it may have on a given simulation.

\subsection{Nonlinear Results} \label{nonlinear}

We now turn to discussion of the nonlinear behavior associated with the various boundary conditions, where unless otherwise stated, results pertain to W7-X geometry under the conditions stated in Section \ref{linear}, with all simulations performed in the gyrofluid approximation.

As mentioned in Section \ref{flux_tube_overview}, fully resolving the dominant $k$-space fluctuation region is required to correctly calculate heat flux values.
The ability of a simulation to satisfy these resolution requirements is directly tied to the radial wavenumber shift of the parallel boundary condition.
If a minimally resolved run requires simulating up to some particular $k_\psi$ value, having a small $k_{\psi}^{shift}$ (based on the boundary condition) clearly forces one to include a large number of simulated modes.
The reverse situation holds true for a large $k_{\psi}^{shift}$.

\begin{figure}[h!]
  \includegraphics[width=3in]{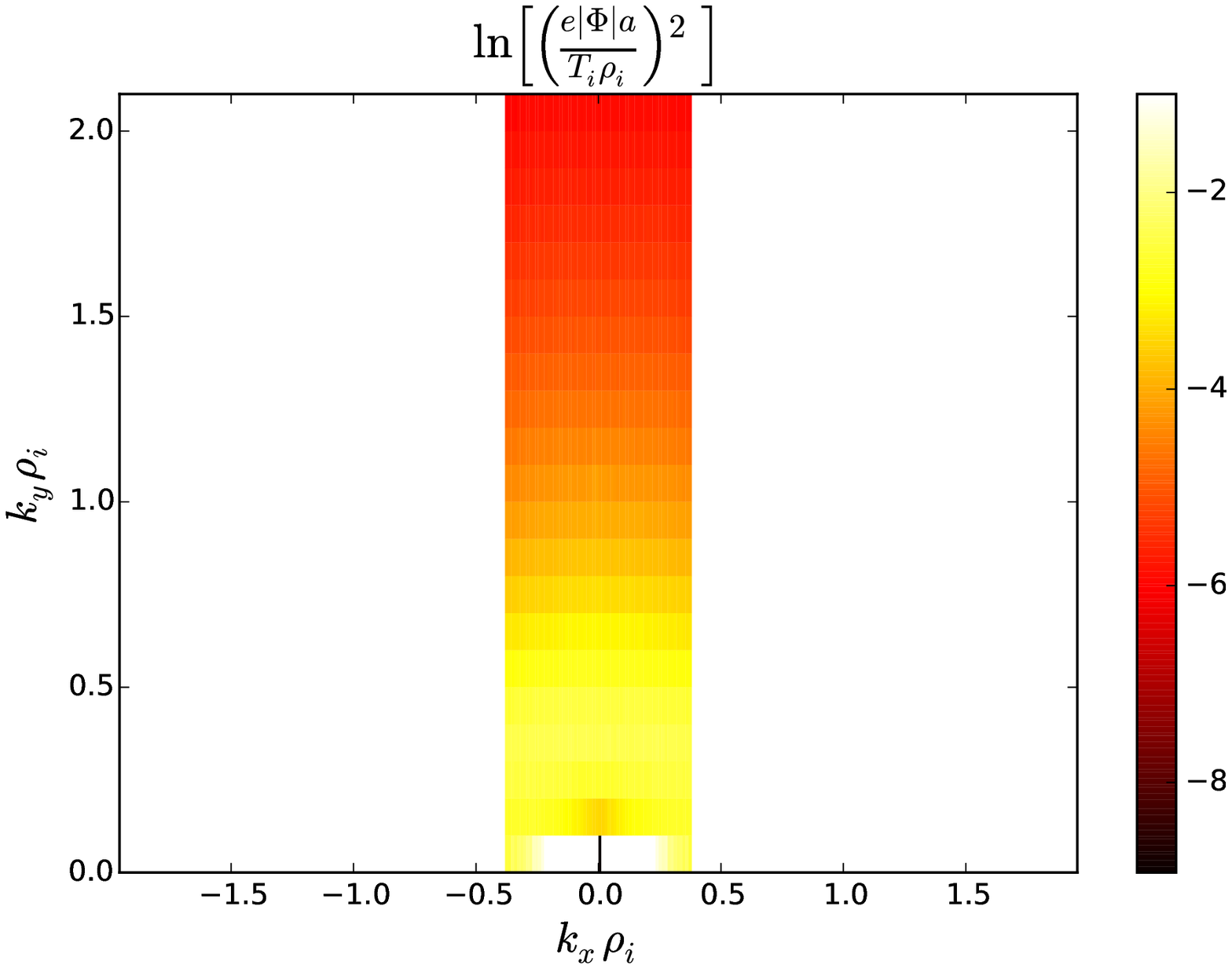}
  \includegraphics[width=3in]{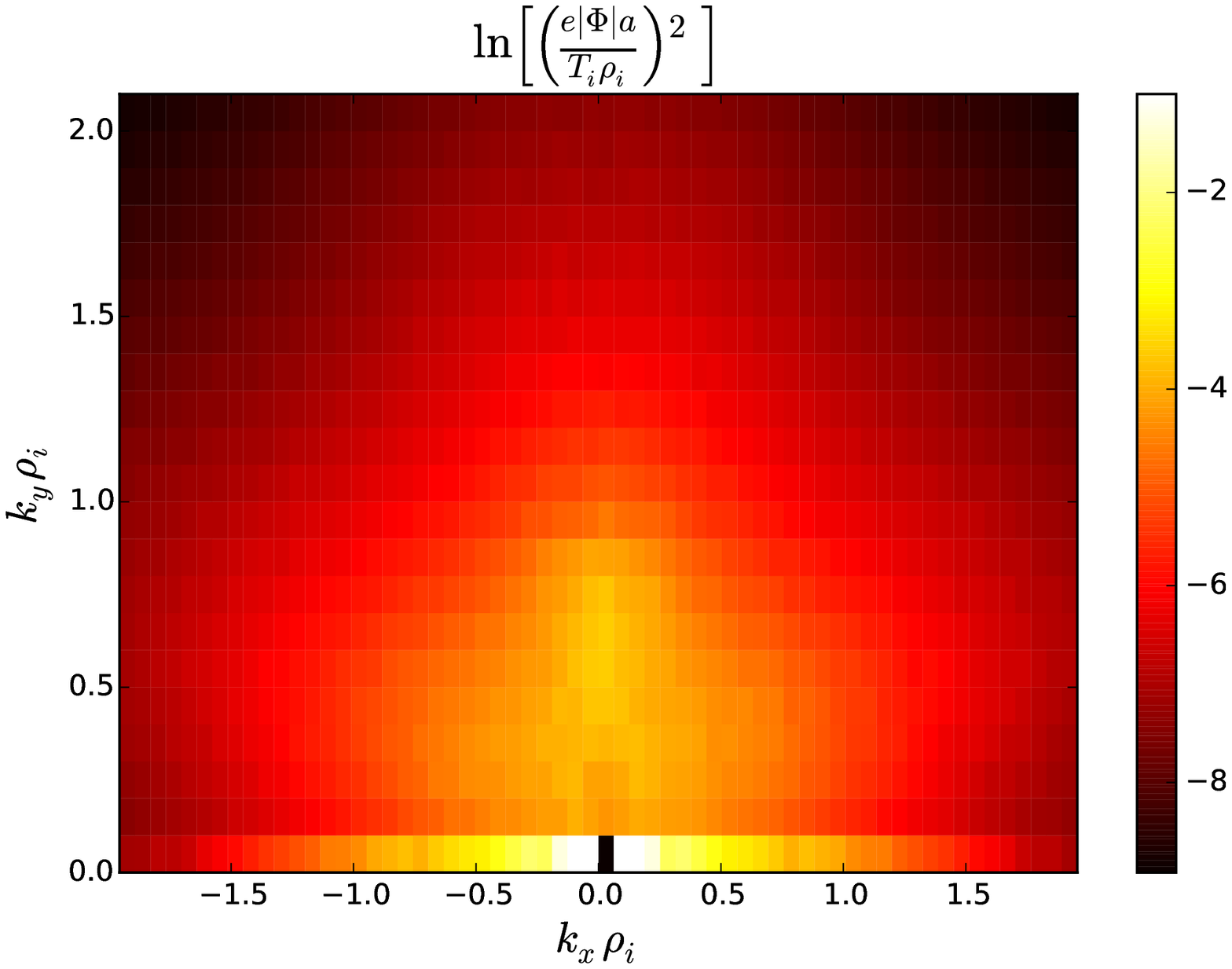}
  \caption{(Color online) - 2D fluctuation spectra with $N_x=96$ using (left) conventional ``twist-and-shift'' covering $k_x=[-0.38,0.38]$ and (right) generalized ``twist-and-shift'' covering $k_x=[-1.96,1.96]$. The increased $k_x$ range in (right) permits fluctuation localization in the domain, while artificially high fluctuations result (left) due to the lack of resolution.}
  \label{fig:spectra}
\end{figure}

Figure \ref{fig:spectra} presents the fluctuation spectrum for a 1 poloidal turn flux tube with $N_x=96$ using the conventional ``twist-and-shift'' boundary condition alongside the spectrum for an unoptimized case of the new boundary condition.
The contrast between the two figures shows unambiguously that a large portion of the fluctuation region is not captured with the conventional boundary condition for this number of radial modes.
The spectrum found with the new boundary condition indicates a localized region of larger relative amplitude in the center, suggesting the simulation contains the most important fluctuations.
On the other hand, the reduced $k_{\psi}$ range in the conventional case does not allow for enough modes to sufficiently capture this region, and the fluctuation amplitudes become artificially large.
Such a case requires one to include more radial modes, and a calculation of transport coefficients at this resolution leads to inaccurate results.

\begin{figure}[h!]
\begin{subfigure}[t]{0.4\textwidth}
\includegraphics[width=\textwidth]{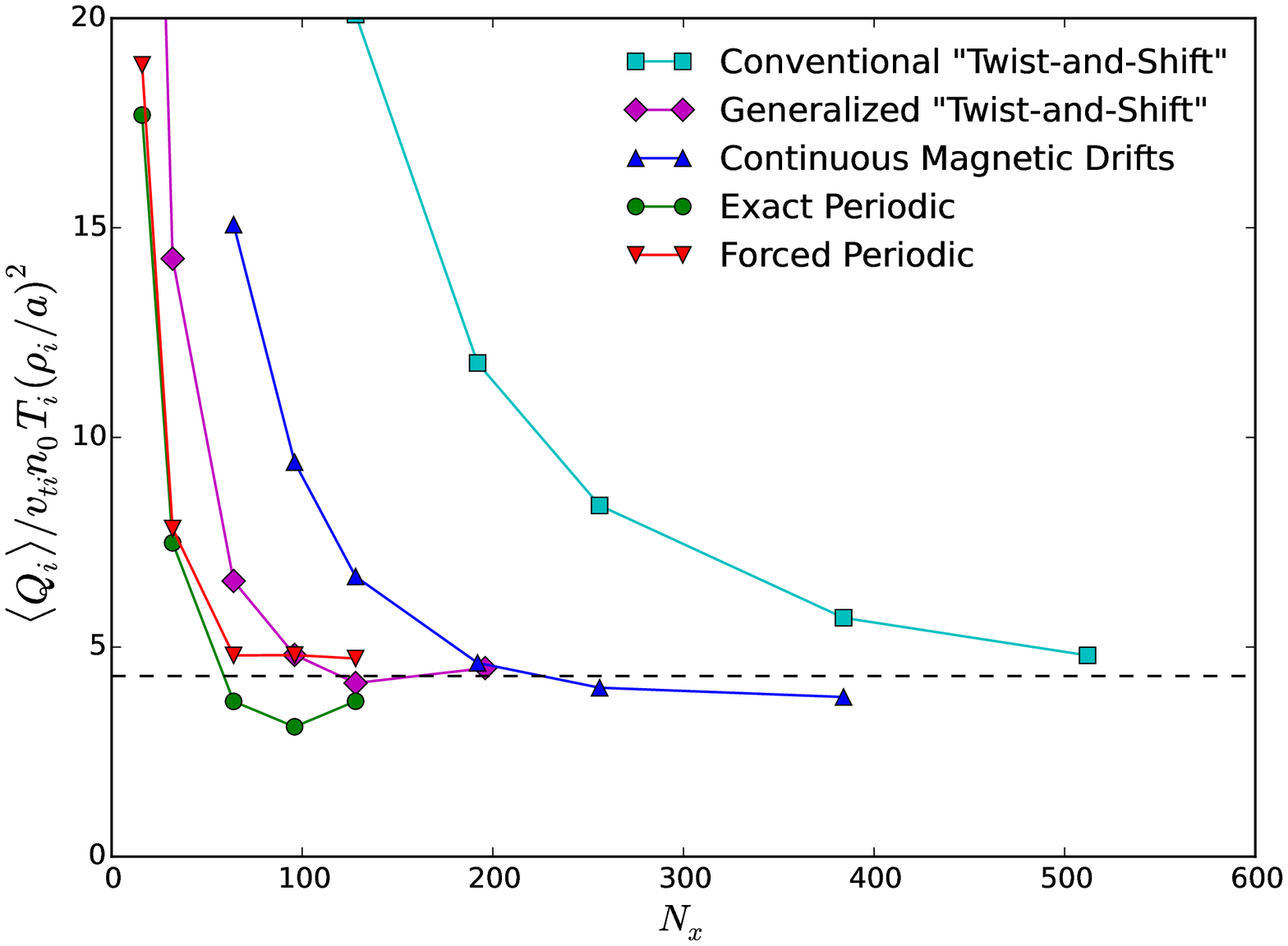}
\caption{}
\label{fig:hflux_w7x}
\end{subfigure}
\begin{subfigure}[t]{0.4\textwidth}
\includegraphics[width=\textwidth]{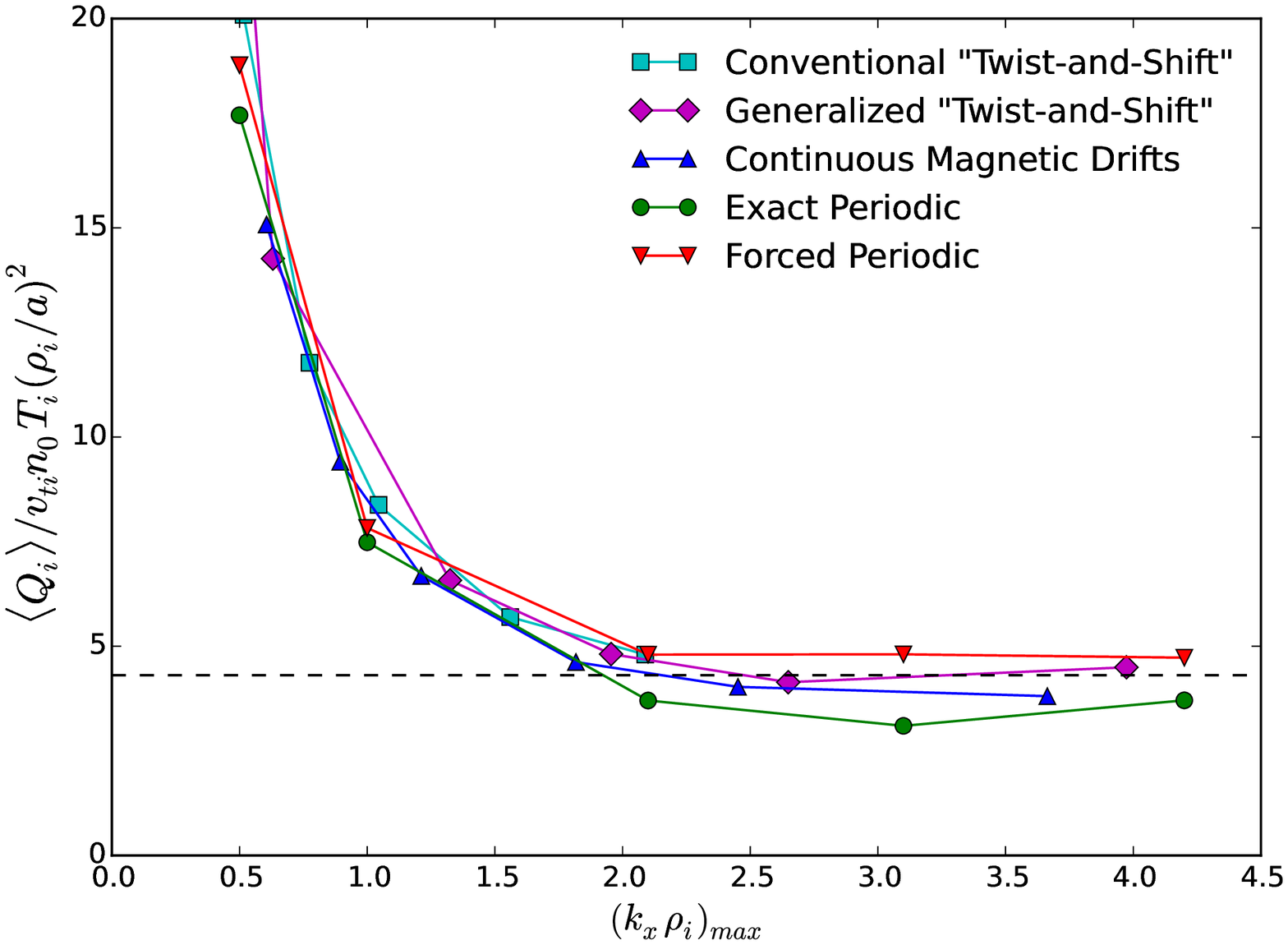}
\caption{}
\label{fig:hflux_kxmax_w7x}
\end{subfigure}
\caption{(Color online) - (a) Saturated heat flux in W7-X as a function of the radial resolution for various boundary condition choices using flux tubes of $\sim$1 poloidal turn. (b) The same data, plotted as a function of the maximum simulated $k_x$ value. Exact flux tube lengths are given in the caption of Figure \ref{fig:gamma_nx}. The dashed line is calculated as the average of the heat flux for the rightmost data point of each boundary condition.}
\end{figure}

The difference in resolution capabilities between the boundary condition is presented in Figure \ref{fig:hflux_w7x} by comparing the saturated heat flux as a function of the number of simulated radial modes, for the boundary condition variations as described in Section \ref{growth_rate}.
Each simulation in Figure \ref{fig:hflux_w7x} uses a flux tube that is $\sim$1 poloidal turn in length, with exact lengths given in the caption of Figure \ref{fig:eig_compare}. 
This figure shows a stark difference in how quickly the results converge with $N_x$ to the correct heat flux (somewhere between 3.5-4.5), based on the chosen boundary condition.
For example, heat flux convergence requires $N_x\simeq96$ for an unoptimized case of the new boundary condition, compared to $N_x\simeq512$ for the conventional boundary condition.
Such a drastic decrease in required resolution leads to a reduction in computational time of $\sim$7x in GryfX.

It should be emphasized here that $k_x^{shift}$ and the domain aspect ratio are directly related, in the sense that larger aspect ratios will produce smaller $k_x^{shift}$ values. 
So while results converged with $N_x\simeq96$ using the unoptimized boundary condition, for a flux tube of 1 poloidal turn $(L_x/L_y=1.59)$, there is no guarantee that every flux tube length will give an aspect ratio in reasonable proximity to 1 (which can be seen in Figure \ref{fig:aspect_gb}).
For a poorly chosen flux tube length with respect to the aspect ratio, it may be that convergence with the new boundary condition is slower with respect to $N_x$, than with conventional ``twist-and-shift" $(L_x/L_y=8.14)$.
However, this also suggests that the required $N_x\simeq192$ for convergence of the continuous magnetic drifts flux tube in Figure \ref{fig:hflux_w7x} is not necessarily directly related to continuity of the magnetic drifts, but may just be a consequence of the domain aspect ratio.  

Another interesting result is the behavior of the two periodic cases (`forced' and `exact').
Surely the most noteworthy outcome pertaining to these runs is the fact that simulations where periodicity is incorrectly enforced converge to the same saturated heat flux as the exact periodicity runs.
This behavior is consistent with the linear calculations in Figure \ref{fig:linear_length} in the sense that the particular choice of boundary condition is irrelevant if the flux tube has sampled ``enough'' of the geometry.
Further, for both the exact and forced periodic simulations, we observe a convergence to the correct heat flux using even fewer radial modes than boundary conditions not employing periodicity.
This provides some evidence that simply applying periodicity may be the optimal choice, even when the theory and continuity properties dictate that it is not strictly valid.

Apart from the differences we find in the required $N_x$ between the boundary condition choices, the previous point made regarding the importance of simulating a large enough region of $k$-space can be further appreciated by plotting the same heat flux data of Figure \ref{fig:hflux_w7x}, but instead as a function of the maximum simulated $k_x$ value in Figure \ref{fig:hflux_kxmax_w7x}.
In doing this, we see that the all heat flux curves nearly overlap, demonstrating that irrespective of which boundary condition is used and how large $N_x$ may need to be, heat flux convergence is ultimately determined by the range of $k$-space that is being simulated.

\begin{figure}[h!]
\includegraphics[width=3in]{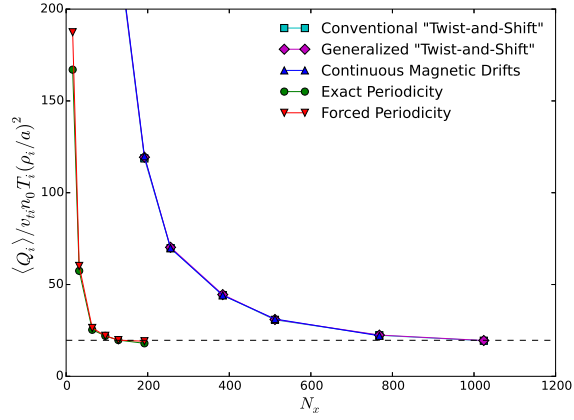}
\caption{(Color online) - Axisymmetric saturated heat flux calculations, where the boundary conditions used in each curve follow Figure \ref{fig:hflux_w7x}, with the following exception: conventional and generalized ``twist-and-shift'', and the flux tube producing continuous magnetic drifts become equivalent in axisymmetry as discussed in Section \ref{axi_limit}. The exact periodic flux tube extends from $[-1.13\pi,1.13\pi]$ to satisfy $[\nabla\psi\cdot\nabla\alpha]_{z_{\pm}} = 0$. The dashed line is calculated as the average of the heat flux for the rightmost data point of each boundary condition.}
\label{fig:hflux_axi}
\end{figure}

While results in Figure \ref{fig:hflux_w7x} are based on a W7-X geometry, one should expect to see consistent behavior in axisymmetry for similar global shear values based on the conditions discussed above that set the resolution requirements.
In Figure \ref{fig:hflux_axi} we present the same study as in Figure \ref{fig:hflux_w7x} for a flux tube in a VMEC-generated axisymmetric geometry, designed to have a global shear value, $\hat{s}=-0.018$, close to that of W7-X.
First of all, the conventional and generalized cases overlap exactly, as one would expect based on how the boundary condition simplifies in axisymmetry.
Beyond this, the simulations model the W7-X case of Figure \ref{fig:hflux_w7x} quite well in reference to the required resolution for convergence to the correct saturated heat flux value.

\begin{figure}[h!]
\includegraphics[width=3in]{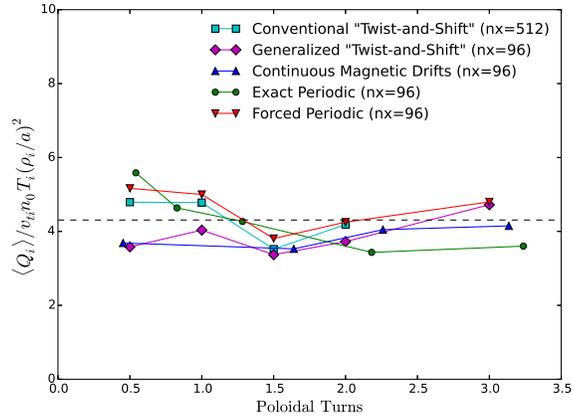}
\caption{(Color online) - Saturated heat flux in W7-X as a function of flux tube length for the various boundary condition choices detailed in Section \ref{growth_rate}. The dashed line is calculated as the average of the heat flux of the rightmost data point for each boundary condition choice in Figure \ref{fig:hflux_w7x}.}
\label{fig:hflux_length}
\end{figure}

Finally, to tie in the linear studies of how flux tube length affects results, the heat flux is calculated for the boundary condition options over a range of flux tube lengths in Figure \ref{fig:hflux_length}.
Within statistical fluctuations, the results appear converged to $\langle\bar{Q}_i\rangle\sim4$ for arguably all flux tube lengths $>0.5$ poloidal turns, which is consistent with the linear growth rate results of Section \ref{linear}.
There is then no reason to expect any significant change in behavior in the limit of the flux tube becoming arbitrarily long.

\section{Conclusions} \label{conclusions}

The ``twist-and-shift'' parallel boundary condition used for gyrokinetic simulations of turbulence in axisymmetric equilibria has been generalized for non-axisymmetric geometries and for configurations with low global magnetic shear.
Twist and shift boundary conditions are associated with field-line-following coordinates in flux tube simulations. 
When the variation of local magnetic shear is strong compared to the global magnetic shear, the flux tube twists and untwists as one moves along the field lines.
Our generalization takes advantage of this phenomenon by using the integrated local magnetic shear to determine the boundary conditions instead of relying only on the global magnetic shear. 
As a result, a considerably smaller periodic computational domain can be identified and additional opportunities for optimization of the simulation domain are exposed. 
The conventional boundary condition of \cite{Beer} is a perfect subset of our generalized formalism. 

Linear stability analyses of W7-X stellarator equilibria have been undertaken using a variety of boundary condition options. 
The growth rates and frequencies are found to be insensitive to the details of the boundary conditions as long as the simulation domain is sufficiently long in the direction of the magnetic field. 
We observe a weak dependence of the calculated eigenvalues on the parallel extent $L$ of the simulation domain as long as $L \gtrsim 2\pi v_i / \omega$. 
This rule of thumb is consistent with Fourier decompositions of the eigenmodes along the field line.
A flux tube that extends at least one poloidal turn was found to be sufficiently long for the W7-X case we examined [see Fig.~(\ref{fig:linear_length})].

In general, nonlinear simulations that are used to estimate turbulent fluxes of heat, {\it etc.,} are very expensive and are the primary targets of our development of improved boundary conditions. 
We have surveyed the behavior of secondary instabilities (which can be highly elongated along the magnetic field) and zonal flows in this context. 
Although we identify cases for which an incorrect (ungeneralized) boundary condition introduces potentially significant parallel discontinuities in the secondary pump waves, we do not observe further serious consequences (such as numerical instability). The importance of this finding will presumably depend on the details of any given numerical discretization.
Zonal flows are strictly periodic along the field line, and are therefore not directly affected by the generalization of the boundary condition. 
Because our approach allows the use of shorter sections of field line, however, we examined the sensitivity of key zonal flow properties to the extent of the flux tube. 
We found that one poloidal turn is evidently long enough to produce consistent short- and long-time zonal flow responses in the W7-X configuration we examined.

The ideal computational domain for a nonlinear problem can be as small as a few correlation lengths in each direction. 
When the global magnetic shear is small, standard ``twist-and-shift'' boundary conditions force one to use a flux tube that can be considerably longer in the radial direction.
Our generalization of the boundary condition makes it possible to hew more closely to the ideal in non-axisymmetric configurations, and we have observed approximately order-of-magnitude speed-ups as a result. 
Once converged, nonlinear heat flux simulations seem to be essentially unaffected by further details of the boundary conditions, 
even as one uses the flexibility enabled by our formalism to satisfy additional continuity properties at the boundaries. 

\begin{acknowledgments}
The authors acknowledge illuminating conversations about this topic with Per Helander, Gabe Plunk, and Ben Faber.
This work was supported by the
U.S. Department of Energy, Office of Science, Office of Fusion Energy Science,
under Award Number DE-FG02-93ER54197.
\end{acknowledgments}

\appendix

\section{Gyrokinetic Equation}
\label{gk_eqn}

In the electrostatic, collisionless limit, the gyrokinetic equation that describes an arbitrary species is given by:
\begin{gather}
  \frac{\partial h}{\partial t} + \bm{v}_{\parallel}\cdot\nabla h + \Big(\langle\bm{v}_{\bm{E}}\rangle + \bm{v}_m\Big)\cdot\nabla_{\perp}h + \langle\bm{v}_{\bm{E}}\rangle\cdot\nabla_{\perp}F_M = \frac{qF_M}{T}\frac{\partial\langle\phi\rangle}{\partial t}.
  \label{eq:gk_eqn}
\end{gather}
The evolved quantity, $h$, represents the non-adiabatic part of the fluctuating distribution function. The quantity $F_M$ represents the background Maxwellian equilibrium distribution.
The angled brackets $\langle...\rangle$ denote a gyroaveraging operation performed at constant particle guiding center, where the quantity $\langle\bm{v}_{\bm{E}}\rangle$ is the gyroaverage of the $\bm{E}\times\bm{B}$ velocity. 
The fluctuating electric field $\bm{E}$ is defined by the gradient of the fluctuating electrostatic potential, $\phi$, which is self-consistently calculated under the assumption of quasineutrality. 
The magnetic drifts are contained in $\bm{v}_m=\bm{\hat{b}}\times\Big[\frac{v_{\perp}^2}{2\Omega}\nabla\log B + \frac{v_{\parallel}^2}{\Omega}(\bm{\hat{b}}\cdot\nabla\bm{\hat{b}})\Big]$, where $\bm{\hat{b}}$ is the unit vector along the magnetic field.

\section{Minimum Flux Tube Length} \label{appendix}

Here, we present an interpretation of the Section \ref{linear} results of linear growth rate convergence with respect to flux tube length. In particular, we attempt to understand the minimum flux tube length for linear convergence (which from Figures \ref{fig:eig_compare_nx7}, \ref{fig:eig_compare}, and \ref{fig:linear_length} appears to be $\sim$1 poloidal turn) using two approaches.

First, one can consider the wave period $2\pi/\omega$
of the $(k_x,k_y)=(0.0,0.2)$ mode where the real frequency is $|\omega|_{k_y=0.2}\approx0.1\; v_{ti}/a$, and estimate the distance $d_i\approx 35\mathrm{m}$ in which a thermal ion travels in that time. This distance can be compared to the length of a 1 poloidal turn flux tube $L_{[-\pi,\pi]}=\int_{-\pi}^{\pi}\mathrm{d}l/(\hat{\bm{b}}\cdot\nabla\theta)$, whereupon one finds that $d_i\sim L_{[-\pi,\pi]}$.
Our numerical results are therefore consistent with the hypothesis that the linear growth rates are converged when the flux tube length is at least $\sim 2\pi v_{ti} / |\omega|$.
Physically, this hypothesis is plausible since thermal particles should sample the correct geometry for the relevant timescale of the mode.

\begin{figure}[h!]
\includegraphics[width=3in]{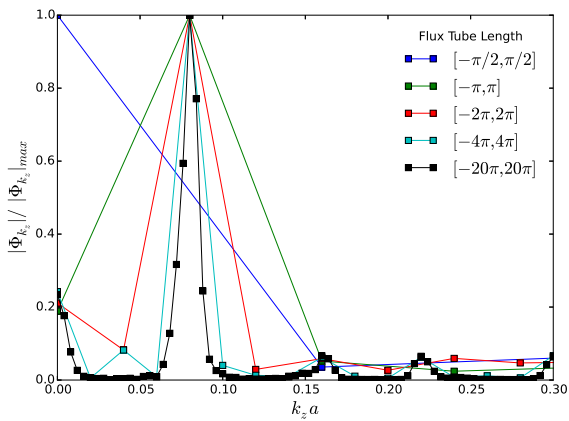}
\caption{(Color online) - Fourier transform of the electrostatic potential along the parallel coordinate for various flux tube lengths, using the Forced Periodic boundary condition.
}
\label{fig:fft_phi}
\end{figure}

A second approach for understanding the minimum flux tube length for convergence is to examine
the mode's parallel structure in Fourier space.
The Fourier transform over the parallel coordinate of $\Phi(z)$ for the $(k_x,k_y)=(0.0,0.2)$ mode has been plotted in Figure \ref{fig:fft_phi} for a range of flux tubes lengths.
It is apparent that the power in $|\Phi_k|$ is strongly peaked around $k_za\approx0.08$, which corresponds to a length of $\sim$1 poloidal turn.
Since results in Figure \ref{fig:linear_length} show that growth rates are mostly converged at 1 poloidal turn for $k_y=0.2$, it appears that linearly, a single data point near $k_za\approx0.08$ is enough to resolve this spike and get the correct growth rate.
Figure \ref{fig:fft_phi} also provides some explanation for the $\sim30\%$ error in growth rates for $\sim$0.5 poloidal turn flux tubes, since power at $k_za\approx0.08$ is binned with $k_za=0$ due to lack of resolution.
The growth rate trend becomes erratic and unreliable as the flux tube length is decreased further, and at small enough lengths the growth rate is quite inaccurate, which can be seen in Figure \ref{fig:linear_length}.
Also in line with resolving the spike at $k_za\approx0.08$ is that as the flux tube is increased past 1 poloidal turn, the growth rate converges to the true result.
We note here that data points in Figure \ref{fig:linear_length} using the conventional ``twist-and-shift" boundary condition for flux tubes with lengths $<0.5$ poloidal turns yield nearly identical results to using the new boundary condition at an unoptimized length.

\end{document}